\documentclass[11pt]{article}
\usepackage[utf8]{inputenc}

\usepackage[pdftex,dvipsnames]{xcolor}  % included in acmart package
\usepackage{amsmath} % included in acmart package
\usepackage{amsthm}
\usepackage{xypic}
\usepackage{array}
\usepackage{stmaryrd}
\usepackage{tikz}
\usepackage{environ} % included in acmart package
\makeatletter
\newsavebox{\measure@tikzpicture}
\NewEnviron{scaletikzpicturetowidth}[1]{%
  \def\tikz@width{#1}%
  \def\tikzscale{1}\begin{lrbox}{\measure@tikzpicture}%
  \BODY
  \end{lrbox}%
  \pgfmathparse{#1/\wd\measure@tikzpicture}%
  \edef\tikzscale{\pgfmathresult}%
  \BODY
}
\makeatother
\usepackage{proof}
\usepackage{mathpartir}
\usepackage[shortlabels]{enumitem}
\usepackage{mathtools}
\usepackage{lscape}
\usepackage{tocloft}
\usepackage{multicol} 
\usepackage{leftidx}
\usepackage{bbm}
\usepackage{rotating}
\usepackage{extarrows}
\usepackage{graphicx} % included in acmart package
\usepackage{placeins}
\usepackage{dsfont}
\usepackage{caption} % included in acmart package

\usepackage{tikz-cd}
\usepackage[margin=1in, top=1in, bottom=1in]{geometry} % included in acmart package
\usepackage{dutchcal}
\usepackage{newpxtext}
\usepackage[varg,bigdelims]{newpxmath}
\usepackage[backend=biber, backref=true, maxbibnames = 10, style = alphabetic]{biblatex}
\usepackage[bookmarks=true, colorlinks=true, linkcolor=blue!50!black,
citecolor=orange!50!black, urlcolor=orange!50!black, pdfencoding=unicode]{hyperref} % included in acmart package
\usepackage[capitalize]{cleveref}
\usepackage{varwidth}
\usepackage{fancyvrb} % fancyhdr included in acmart package
\usepackage{xcolor} % included in acmart package

\usepackage{listings}

\lstdefinelanguage[2026]{CaMPL}{%
  keywords={=},%
  alsoletter={=},
  otherkeywords={:=,::,|,=>,->},%
  morekeywords={[1]data,codata,protocol,coprotocol,proc,%
    do,%
    fun,
    case,fold,unfold,switch,defn,where,if,then,else,
	import
  },%
  morekeywords={[2]Console,StringTerminal,Timer,IntTerminal,CharTerminal,Terminal,IntConsole,CharConsole},
  morekeywords={[3]Put,Get,TopBot,Store},
  morekeywords={[4]hput,hcase,put,get,halt,plug,close,fork,split,(*),(+),store,use,race,neg},
  sensitive,%
  morecomment=[l]--,%
  morecomment=[n]{\{-}{-\}},%
  morestring=[b]"%
}[keywords,comments,strings]%
\definecolor{dkgreen}{rgb}{0,0.6,0}
\definecolor{gray}{rgb}{0.5,0.5,0.5}
\definecolor{purple}{rgb}{0.8,0,0.3}
\definecolor{orange}{rgb}{1,0.4,0}
\definecolor{lightlightgray}{rgb}{.95,.95,.95}
\definecolor{lightgray}{rgb}{.9,.9,.9}
\definecolor{lightgray2}{rgb}{.85,.85,.85}
\definecolor{darkgray}{rgb}{.4,.4,.4}
\definecolor{darkred}{rgb}{0.6,0,0}
\definecolor{macyellow}{rgb}{0.980,0.804,0.353}

\lstdefinestyle{CaMPLstyle}{%
  language=CaMPL,%
  keywordstyle={[1]\color{darkred}},%
  keywordstyle={[2]\color{blue}},%
  keywordstyle={[3]\color{dkgreen}},%
  keywordstyle={[4]\color{purple}},%
  commentstyle={\itshape\color{orange}},%
  identifierstyle=\color{black},%
  stringstyle=\color{teal},%
  showstringspaces=false,%
  basicstyle=\ttfamily\small,%
  tabsize=2,%
  showtabs=false,%
  tab={\color{gray}\rightarrowfill},%
  numbers=left,%
  numberstyle=\color{gray},%
  breaklines=true,%
  frame=single,%
  xleftmargin=2em,% <--- ADD THIS HERE
  numbersep=8pt,%  <--- ADD THIS HERE
  rulecolor=\color{lightgray2},%
  rulesepcolor=\color{gray},%
  backgroundcolor=\color{lightgray},%
}%
\usepackage{todonotes}

\definecolor{darkMagenta}{HTML}{DE3163}
\newcommand{\tc}[1]{\textcolor{darkMagenta}{\textit{#1}}}

\usepackage{todonotes}

\numberwithin{dummy}{section}

\theoremstyle{definition}

\theoremstyle{definition}

\numberwithin{equation}{section}

\newcommand{\tri
}{\triangleleft
}

\newcommand{\X}{\mathbb{X}}
\newcommand{\A}{\mathbb{A}}

\newcommand{\ox}{\otimes}

\newcommand{\oa}{\oplus}

\newlength{\llcfoo}



\makeatletter

\newdimen\w@dth

\def\setw@dth#1#2{\setbox\z@\hbox{\scriptsize $#1$}\w@dth=\wd\z@
\setbox\@ne\hbox{\scriptsize $#2$}\ifnum\w@dth<\wd\@ne \w@dth=\wd\@ne \fi
\advance\w@dth by 1.2em}

\def\t@^#1_#2{\allowbreak\def\n@one{#1}\def\n@two{#2}\mathrel
{\setw@dth{#1}{#2}
\mathop{\hbox to \w@dth{\rightarrowfill}}\limits
\ifx\n@one\empty\else ^{\box\z@}\fi
\ifx\n@two\empty\else _{\box\@ne}\fi}}
\def\t@@^#1{\@ifnextchar_ {\t@^{#1}}{\t@^{#1}_{}}}

\def\t@left^#1_#2{\def\n@one{#1}\def\n@two{#2}\mathrel{\setw@dth{#1}{#2}
\mathop{\hbox to \w@dth{\leftarrowfill}}\limits
\ifx\n@one\empty\else ^{\box\z@}\fi
\ifx\n@two\empty\else _{\box\@ne}\fi}}
\def\t@@left^#1{\@ifnextchar_ {\t@left^{#1}}{\t@left^{#1}_{}}}

\def\two@^#1_#2{\def\n@one{#1}\def\n@two{#2}\mathrel{\setw@dth{#1}{#2}
\mathop{\vcenter{\hbox to \w@dth{\rightarrowfill}\kern-1.7ex
                 \hbox to \w@dth{\rightarrowfill}}%
       }\limits
\ifx\n@one\empty\else ^{\box\z@}\fi
\ifx\n@two\empty\else _{\box\@ne}\fi}}
\def\tw@@^#1{\@ifnextchar_ {\two@^{#1}}{\two@^{#1}_{}}}

\def\tofr@^#1_#2{\def\n@one{#1}\def\n@two{#2}\mathrel{\setw@dth{#1}{#2}
\mathop{\vcenter{\hbox to \w@dth{\rightarrowfill}\kern-1.7ex
                 \hbox to \w@dth{\leftarrowfill}}%
       }\limits
\ifx\n@one\empty\else ^{\box\z@}\fi
\ifx\n@two\empty\else _{\box\@ne}\fi}}
\def\t@fr@^#1{\@ifnextchar_ {\tofr@^{#1}}{\tofr@^{#1}_{}}}

\newdimen\W@dth
\def\setW@dth#1#2{\setbox\z@\hbox{$#1$}\W@dth=\wd\z@
\setbox\@ne\hbox{$#2$}\ifnum\W@dth<\wd\@ne \W@dth=\wd\@ne \fi
\advance\W@dth by 1.2em}

\def\T@^#1_#2{\allowbreak\def\N@one{#1}\def\N@two{#2}\mathrel
{\setW@dth{#1}{#2}
\mathop{\hbox to \W@dth{\rightarrowfill}}\limits
\ifx\N@one\empty\else ^{\box\z@}\fi
\ifx\N@two\empty\else _{\box\@ne}\fi}}
\def\T@@^#1{\@ifnextchar_ {\T@^{#1}}{\T@^{#1}_{}}}

\def\T@left^#1_#2{\def\N@one{#1}\def\N@two{#2}\mathrel{\setW@dth{#1}{#2}
\mathop{\hbox to \W@dth{\leftarrowfill}}\limits
\ifx\N@one\empty\else ^{\box\z@}\fi
\ifx\N@two\empty\else _{\box\@ne}\fi}}
\def\T@@left^#1{\@ifnextchar_ {\T@left^{#1}}{\T@left^{#1}_{}}}

\def\Tofr@^#1_#2{\def\N@one{#1}\def\N@two{#2}\mathrel{\setW@dth{#1}{#2}
\mathop{\vcenter{\hbox to \W@dth{\rightarrowfill}\kern-1.7ex
                 \hbox to \W@dth{\leftarrowfill}}%
       }\limits
\ifx\N@one\empty\else ^{\box\z@}\fi
\ifx\N@two\empty\else _{\box\@ne}\fi}}
\def\T@fr@^#1{\@ifnextchar_ {\Tofr@^{#1}}{\Tofr@^{#1}_{}}}

\def\Two@^#1_#2{\def\N@one{#1}\def\N@two{#2}\mathrel{\setW@dth{#1}{#2}
\mathop{\vcenter{\hbox to \W@dth{\rightarrowfill}\kern-1.7ex
                 \hbox to \W@dth{\rightarrowfill}}%
       }\limits
\ifx\N@one\empty\else ^{\box\z@}\fi
\ifx\N@two\empty\else _{\box\@ne}\fi}}
\def\Tw@@^#1{\@ifnextchar_ {\Two@^{#1}}{\Two@^{#1}_{}}}

\def\to{\@ifnextchar^ {\t@@}{\t@@^{}}}
\def\from{\@ifnextchar^ {\t@@left}{\t@@left^{}}}
\def\tofro{\@ifnextchar^ {\t@fr@}{\t@fr@^{}}}
\def\To{\@ifnextchar^ {\T@@}{\T@@^{}}}
\def\From{\@ifnextchar^ {\T@@left}{\T@@left^{}}}
\def\Two{\@ifnextchar^ {\Tw@@}{\Tw@@^{}}}
\def\Tofro{\@ifnextchar^ {\T@fr@}{\T@fr@^{}}}

\makeatother

\tikzstyle{strings}=[baseline={([yshift=-.5ex]current bounding box.center)}]

\tikzset{every picture/.append style={scale=.5}, transform shape, strings}

\tikzset{%
symbol/.style={%
draw=none,
every to/.append style={%
edge node={node [sloped, allow upside down, auto=false]{$#1$}}}
}
}

\usetikzlibrary{shapes.geometric}
\usetikzlibrary{patterns}
\usetikzlibrary{fit}
\usetikzlibrary{positioning}
\usetikzlibrary{calc}
\usetikzlibrary{arrows}
\usetikzlibrary{decorations.markings}
\usetikzlibrary{decorations.pathreplacing}
\usetikzlibrary{shapes}

\pgfdeclarelayer{nodelayer}
\pgfdeclarelayer{edgelayer}
\pgfsetlayers{edgelayer,nodelayer,main}

\tikzset{simple/.style={}}
\tikzset{nothing/.style={outer sep=-3.4pt}}

\tikzset{map/.style={draw,fill=white, thick, rectangle}}
\tikzset{mapblack/.style={draw,fill=black, rectangle}}

\tikzstyle{filled}=[-, fill=black]

\tikzset{dot/.style={thick, fill=black, circle, scale=1, inner sep = .05cm}}

\tikzset{oa/.style={draw, scale=0.9,minimum height=.1cm,circle,append after command={
[shorten >=\pgflinewidth, shorten <=\pgflinewidth,]
(\tikzlastnode.north) edge (\tikzlastnode.south)
(\tikzlastnode.east) edge (\tikzlastnode.west)
} } }

\tikzset{ox/.style={draw, scale=0.9,minimum height=.1cm,circle,append after command={
[shorten >=\pgflinewidth, shorten <=\pgflinewidth,]
(\tikzlastnode.north west) edge (\tikzlastnode.south east)
(\tikzlastnode.north east) edge (\tikzlastnode.south west) } } }

\tikzset{coprod/.style={draw, thick, scale=0.75, circle } }

\tikzset{prod/.style={draw, fill=black, scale=0.75, circle } }

\tikzset{circ/.style={
shape=circle, inner sep=1pt, draw}}

\tikzstyle{none}=[inner sep=-1pt]
\tikzstyle{circle}=[shape=circle,draw]

\tikzstyle{onehalfcircle}=[shape=circle, scale=1.5, draw]
\tikzstyle{twocircle}=[shape=circle, scale=2, draw]
\tikzstyle{black}=[shape=circle, fill=black, draw]

\tikzstyle{head}=[fill=white, draw=black, shape=circle, scale=2, thick]
\tikzstyle{grayhead}=[dashed, fill=white, draw=gray, shape=circle, scale=2]
\tikzstyle{process}=[fill=white, draw=black, shape=circle, scale=2, thick]

\tikzset{wires/.style={}}

\tikzset{box/.style={inner sep=0pt, thick, draw=black, text height=1.5ex, text depth=.25ex, 
text centered, minimum height=3em, anchor=center}}

\tikzcdset{every label/.append style = {font = \Large}}

\newcommand{\linmonw} {\xymatrixcolsep{4mm} \xymatrix{ \ar@{-||}[r]^{\circ} & }}
\newcommand{\linmonwl} {\xymatrixcolsep{4mm} \xymatrix{ \ar@{-||}[r]^{\otimes\;\tri} & }}
\newcommand{\linmonwr} {\xymatrixcolsep{4mm} \xymatrix{ \ar@{-||}[r]^{\tri\;\otimes} & }}
\newcommand{\linmonwrdavid} {\xymatrixcolsep{4mm} \xymatrix{ \ar@{-||}[r]^{\otimes\;\tri} & }}
\newcommand{\linmonwldavid} {\xymatrixcolsep{4mm} \xymatrix{ \ar@{-||}[r]^{\tri\;\otimes} & }}
\newcommand{\linmondavid} {\xymatrixcolsep{4mm} \xymatrix{ \ar@{-||}[r] & }}

\newcommand{\lincomonb} {\xymatrixcolsep{4mm} \xymatrix{ \ar@{-||}[r]_{\bullet} & }}
\newcommand{\lincomonw} {\xymatrixcolsep{4mm} \xymatrix{ \ar@{-||}[r]_{\circ} & }}
\newcommand{\lincomonwr} {\xymatrixcolsep{4mm} \xymatrix{ \ar@{-||}[r]_{\tri\;\otimes} & }}
\newcommand{\lincomonwl} {\xymatrixcolsep{4mm} \xymatrix{ \ar@{-||}[r]_{\otimes\;\tri} & }}

\newcommand{\linbialgw} {\xymatrixcolsep{4mm} \xymatrix{ \ar@{-||}[r]^{\circ}_{\circ} & }}
\newcommand{\linbialgwl} {\xymatrixcolsep{4mm} \xymatrix{ \ar@{-||}[r]^{\otimes\;\tri}_{\otimes\;\tri} & }}
\newcommand{\linbialgwr} {\xymatrixcolsep{4mm} \xymatrix{ \ar@{-||}[r]^{\tri\;\otimes}_{\tri\;\otimes} & }}
\newcommand{\linbialgwb} {\xymatrixcolsep{4mm} \xymatrix{ \ar@{-||}[r]^{\circ}_{\bullet} & }}

\newcommand{\blackman}{
\begin{tikzpicture}[scale=1.5]
	\begin{pgfonlayer}{nodelayer}
		\node [style=head] (0) at (-3, 5) {};
		\node [style=none] (1) at (-3, 4) {};
		\node [style=none] (2) at (-3.25, 3.75) {};
		\node [style=none] (3) at (-2.75, 3.75) {};
		\node [style=none] (4) at (-3.25, 4.45) {};
		\node [style=none] (5) at (-2.75, 4.45) {};
	\end{pgfonlayer}
	\begin{pgfonlayer}{edgelayer}
		\draw[thick] (0) to (1.center);
		\draw[thick] (1.center) to (2.center);
		\draw[thick] (1.center) to (3.center);
		\draw[thick] (4.center) to (5.center);
	\end{pgfonlayer}
\end{tikzpicture} }

\newcommand{\biglens}[2]{
     \begin{bmatrix}{\vphantom{f_f^f}#2} \\ {\vphantom{f_f^f}#1} \end{bmatrix}
}
\newcommand{\littlelens}[2]{
     \begin{bsmallmatrix}{\vphantom{f}#2} \\ {\vphantom{f}#1} \end{bsmallmatrix}
}
\newcommand{\coclose}[2]{
  \relax\if@display
     \biglens{#2}{#1}
  \else
     \littlelens{#2}{#1}
  \fi
}

\title{Categorical Message Passing Language (CaMPL): \newline Syntax and Semantics}
\author{
    Robin Cockett\thanks{University of Calgary} \and Daniel Kiyoshi Hashimoto\thanks{Universidade Federal do Rio de Janeiro} \and Alexanna {Little Berg}* \and Priyaa Varshinee Srinivasan\thanks{Tallinn University of Technology, Estonia. This work was co-funded by the European Union and Estonian Research Council through the Mobilitas 3.0
(MOB3JD1227).}}
\date{\today}

\begin{document}

\maketitle

\begin{abstract}
We introduce a novel functional-style concurrent programming language called Categorical Message Passing Language (CaMPL) which is designed using the mathematics of linear actegories. 
This mathematical underpinning gives CaMPL programs useful properties such as deadlock freedom, and additionally, livelock freedom for programs without general recursive processes.

We explore CaMPL's type system through a series of code examples. 
The current proto-alpha version of the compiler and the abstract machine is implemented in Haskell. 
A reader is encouraged to experiment with writing CaMPL programs either using the online compiler  \url{https://campl-app.vercel.app/} or by installing CaMPL from \url{https://campl-ucalgary.github.io} -- our website has detailed instructions on how to run CaMPL code.
%Categorical Message Passing Language, to the best of our knowledge, the only\footnote{Is this true?} concurrent programming language with mathematical underpinnings. 
%It is to be considered as a proof-of-concept of a programming language.
\end{abstract}

\tableofcontents

\section{Introduction}
%
%Explain current state of concurrency and why CaMPL is novel (Curry-Howard-Lambek-like correspondence), and its history.
%
%This section should introduce the salient details of CaMPL (like it should be a highlight reel of CaMPL. if people don't read past this section they should still be able to be like ``oh yeah i heard of CaMPL it was cool because of x, y, z'') which could include some of the following points:

Categorical Message Passing Language (CaMPL) is a functional-style concurrent programming language that uses asynchronous message-passing concurrency.
It implements a type system based on linear logic that was described in Cockett and Pastro’s paper ``The Logic of Message Passing'' \cite{CoP07}. The primary result of their paper was defining and proving a concurrent analogue of the functional programming Curry-Howard-Lambek correspondence or proofs-as-programs principle \cite{Howard1980, Lambek1969, Lambek1972}. 
Their paper diverges from the approach taken by Abramsky et al. in 
\cite{ABRAMSKY19933, Abr1993IntCats, Abramsky1999ConcurrentGA, AGN2000} which used exclusively the types and rules of linear logic that were introduced by Girard in \cite{GIRARD19871}.
It also diverges from the popular approach of using linear logic to retrospectively develop a type system for the \(\pi\)-calculus (comprehensively presented by Milner in \cite{Milner1993TheP}) which has produced a significant body of work especially after the connection to session types was established \cite{Honda1993TypesFD, BELLIN199411, Kob96, Barber1997Action, CPf2010, Wa12, CairesPT16, KMP19, AG19, QKB21}.
Instead, Cockett and Pastro introduced a two-tiered logic that augments linear logic with new rules for passing messages of sequential data, such as strings or integers.
Consequently, CaMPL is a two-tiered programming language.
We motivate this design decision with the following excerpt from Cockett and Pastro’s paper: 
%consisting of a sequential part, called the message logic, that interacts with a concurrent part, called the message-passing logic. 
%In the message-passing logic, concurrent processes pass messages of sequential data back and forth along communication channels.
%% do we want to compare message passing concurrency to shared memory and mention the benefits in terms of distributed systems that cannot share memory?? (even though the current implementation still heavily relies on the processes sharing memory)
\begin{quote}
We model message passing using a two tier logic. There is a logic for the messages whose proofs should be thought of as ordinary sequential programs.
Then there is a logic of message passing which is built on top of the logic of messages. The two logics are really quite distinct: the message logic is concerned with what we classically view as computation, while the second logic is concerned with manipulating the channels of communication.

[...] However, even the briefest perusal of the rules, indicates that the logic concerned with managing channels is at least as complicated as the sequential programming logic. Furthermore, there are quite significant
interactions between the two levels. 
[...]
 Thus, we believe programming language designers should be thinking in terms of developing integrated two tier languages in order to give high-level support for concurrent programming [...]. 
\end{quote}
The categorical semantics Cockett and Pastro give for the two-tiered logic is a {\em linear actegory}.
% in which the category of messages and sequential functions acts on the category of communication channels and concurrent processes in two directions.
%The covariant action passes messages in the forward direction, which we describe in this article as ``left to right,'' and the contravariant action passes messages in the backward direction, which we describe as ``right to left.''
%CaMPL programs are composed of multiple concurrent processes which communicate with each other by passing messages along communication channels.
%At any instance in run-time, the topology of a CaMPL program is a finite acyclic graph consisting of processes which are nodes of the graph and channels which are the edges of the graph.
% The property of acyclicity at run-time is guaranteed by CaMPL's type system.
A linear actegory minimally consists of a monoidal category acting covariantly and contravariantly on a linearly distributive category, 
%(LDC), 
satisfying certain coherences~\cite{CoP07, CS97}. 
%Linearly distributive categories are the categorical semantics of multiplicative linear logic~\cite{CS97}. 

The concurrent type system provided by the message-passing logic ensures that processes can never be connected in a cycle.
That is, at any instance in run-time, the topology of a program is guaranteed to be a finite acyclic graph consisting of processes as nodes and channels as edges.
%This property comes from the interaction between composition and the non-isomorphism linear distributors in the LDC.
The benefit of this property is that problems caused by cycles, such as deadlocks and livelocks, do not occur \cite{Ly18}.
However, similar to proving termination in the sequential case, livelock freedom is only guaranteed if we disallow general recursion.
Deadlock freedom is still guaranteed, as is common in concurrent type systems based on linear logic \cite{BELLIN199411, CPf2010, Wa12, CairesPT16, QKB21}.

% how the language implementation started;  history of work on campl prashant thesis, jared thesis, xanna thesis, melika thesis
The first implementation of the CaMPL compiler was written by Kumar as described in \cite{Ku18}.
This version implemented sequential and concurrent type systems.
It added features to define custom concurrent data types called protocols and coprotocols.
The categorical semantics for protocols and coprotocols were given by Yeasin in \cite{Ye12}.
This compiler performed both a type inference and type checking.
The second implementation of the compiler was written by Pon as described in \cite{Po21, Po22}.
This version re-implemented the existing features and added controlled non-determinism in the form of ``races.''
Changes to the categorical semantics to include non-determinism were outlined by Little in \cite{xanna, xanna2}.
The most recent feature added to the compiler was higher-order message passing which allows processes to send messages that contain encoded processes which can be decoded and invoked when they are received.
The implementation of this feature and the changes to the categorical semantics were written by Norouzbeygi in \cite{No25, Cockett_2026}. 
The current implementation of CaMPL is available at \url{https://campl-ucalgary.github.io/}.

In this article, we illustrate CaMPL's two-tiered type system and demonstrate the above features through a series of code examples.
The brief overview in Section~\ref{Sec:brief_overview} considers two basic examples without explaining specific details.
We discuss the concurrent tier of CaMPL, which corresponds to the message-passing logic, in Section~\ref{Sec:ConcurrentTypes}.
We discuss the sequential tier of CaMPL, which corresponds to the message logic, in Section~\ref{Sec:SequentialTypes}.
We will also discuss how the value of sequential data can be used to control a process's execution.
The key concepts of the above programming features will be covered by these two sections, and the corresponding channel types and process commands are summarized in Table~\ref{Table:summary}.
If one is interested in writing CaMPL programs, we provide finer details in Section~\ref{sec:concurrent-operators}.
We also cover some more complicated examples.
Finally, in Section~\ref{Sec:math_underp}, we briefly discuss  the categorical and logical semantics described by Cockett and Pastro and explain how CaMPL emerged from them.
In Appendix~\ref{sec:appendix}, we provide examples of complete programs that were written by collecting code from the examples throughout this article.

% from all my reading about session types i have learned some things.
% exponential !A expresses capability to receive from an unbounded number of concurrent processes (like a server) but shouldn't it also be able to send because A is a session type?
% also can the processes connected to the server also be connected to each other??
% with only this we still have deadlock freedom, but there is another:
% exponential ?A expresses capability to send to an unbounded number of concurrent processes (like a broadcast)
% i don't know if this will still have deadlock freedom. 
% BUT could this help with fault tolerance? like it could facilitate controlled redundancy?
% semantics could be like the chemical concurrency stuff that's non-deterministic i think?

\section{A brief overview of CaMPL}
\label{Sec:brief_overview}

As is tradition, we will introduce CaMPL with a ``Hello World!'' example program, shown in Example \ref{ex:helloworld}.
We use \tc{this style} whenever we introduce CaMPL terminology.
We demonstrate a \tc{process} named \lstinline$helloworld$ printing \lstinline$"Hello World!"$ by sending the string as a message on a \tc{channel} named \lstinline$console$.
The \lstinline$console$ channel has type \lstinline$Console$ which is a special channel type built-in to the compiler to connect any CaMPL program to the terminal from which it is run.
\begin{lstlisting}[label=ex:helloworld, caption=Hello World!, name=helloworld] 
proc helloworld :: | Console => = 
	| console => -> do
		hput ConsolePut on console
		put "Hello World!" on console    -- sends message to the console        
		hput ConsoleClose on console
		halt console                     -- closes console channel and halts               
\end{lstlisting}
The \lstinline$helloworld$ process can be invoked by calling it in the main process, \lstinline$run$, and giving it the \lstinline$console$ channel as follows:
\begin{lstlisting}[name=helloworld]
proc run = 
	| console => -> helloworld( | console => )  -- creates helloworld process
\end{lstlisting}

In Example \ref{ex:serverecho}, we will demonstrate communication between two user-defined processes: \lstinline$client$ and \lstinline$server$. 
The \lstinline$client$ will send a string to \lstinline$server$, and \lstinline$server$ will receive it and echo it back.
Then, \lstinline$client$ and \lstinline$server$ will halt.
The main process will create the \lstinline$client$ and \lstinline$server$ processes, and it will use a \tc{process command} called \lstinline$plug$ to connect them by a channel named \lstinline$ch$.

\begin{lstlisting}[label=ex:serverecho, caption=Server echos Client's message]
 proc run = 
 	| => -> plug						-- connects processes by shared channel ch
 		client( | => ch )			-- creates client process
 		server( | ch => )			-- creates server process
\end{lstlisting}
The CaMPL compiler will infer the \tc{channel type} of \lstinline$ch$ using the process' definitions. 
%We will discuss channel types in Section~\ref{Sec:channeltypes}.

The client process has an \tc{output polarity channel} \lstinline$ch$ which connects it to the server.
%We discuss polarity in Section~\ref{Sec:polarities}.
It sends its message with \lstinline$put$, and it receives the server's echo with \lstinline$get$.
Then, it uses \lstinline$halt$ to close the channel with the server and halt.
We can define \lstinline$client$ as follows:
\begin{lstlisting}
proc client = 
	| => ch ->							-- ch is an output polarity channel
		on ch do 
			put "Hello Server!"	-- sends message to the server      
			get echo						-- receives server's echo  
			halt
\end{lstlisting}

The server process has an \tc{input polarity channel} \lstinline$ch$ which connects it to the client.
It listens to the client and receives the client's message with \lstinline$get$, echos the message back with \lstinline$put$, and finally closes the channel and halts with \lstinline$halt$.
We can define \lstinline$server$ as follows:
\begin{lstlisting}
proc server =  
	| ch => ->							-- ch is an input polarity channel
		on ch do
			get msg							-- receives client's message
			put msg							-- sends message back
			halt
\end{lstlisting}	

These simple programs conceptually exemplify how CaMPL programs are written.
In fact, the lines of the code in the second program are out of order as the \lstinline$run$ process should be the final process defined in a program.
We give a full program with the lines in order in Example~\ref{ex:serverechofull}.
Furthermore, the runnable program obtained by reordering the lines still would not produce any observable effects since none of the processes are connected to the outside world.

A reader should be left with open questions such as ``what was the type of channel \lstinline$ch$?'' ``what is polarity?'' and ``what else can this language do?''
We hope that these questions motivate the reader to continue on to the more complicated examples in the following sections.

\section{Concurrent tier}
\label{Sec:ConcurrentTypes}

The concurrent tier of CaMPL is concerned with managing interactions between processes along channels.
At any instance in run-time, the topology of a program is a finite acyclic graph consisting of processes as nodes and channels as edges. 
The property of acyclicity is guaranteed by the type system, and it ensures that programs will never deadlock.
Furthermore, a program without any general recursive processes will never deadlock or livelock.

The CaMPL compiler performs both a type inference and type check at compile-time to ensure the compiled program follows the type system.
This section highlights the features of CaMPL that constitute its concurrent type system.
We synthesize content from \cite{Ku18, Po21, Po22, No25}.

\subsection{Processes}
\label{Sec:Processes}

Processes are the main actors of a CaMPL program.
A process is specified by the keyword \lstinline$proc$ followed by a process name, an optional type signature, and lists of its variables and channels.

In Example \ref{ex:helloworld}, the type signature of process \lstinline$helloworld$ was "\lstinline$ :: | Console => $" which indicates that it has access to one channel of type \lstinline$Console$.
This channel is given the name \lstinline$console$ as indicated by "\lstinline$ = | console => $" on the next line.
We will discuss type signatures in Section~\ref{Sec:channeltypes}.

The final process defined in a program, in which the program's execution begins, must have the name \lstinline$run$. 
This process is the main process and the only process which can create \tc{service channels}, such as the \lstinline$Console$, for communication with the outside world.
We will discuss service channels in more detail in Section~\ref{Sec:ServiceChannels}.

For now, we will consider the definition of process \lstinline$broadcast$ in Example \ref{ex:broadcastmsg}:
\begin{lstlisting}[label=ex:broadcastmsg, caption=Process that broadcasts a message from a single source] 
proc broadcast = 
	ack_msg | source => dest1, dest2 -> do
		get msg on source								-- receive message from source
		put msg on dest1								-- broadcast message to other processes
		put msg on dest2
		put ack_msg on source						-- send acknowledgement message to source
		close source										-- close all channels and halt
		close dest1
		halt dest2  -- the last channel is closed with the halt command
\end{lstlisting}
Observe that line 2 consists of three comma-seperated lists: 1)~variables, e.g. \lstinline$ack_msg$, 2)~input polarity channels, e.g. \lstinline$source$, and 3)~output polarity channels, e.g. \lstinline$dest1,  dest2$.
\tc{Variables} are instances of sequential types and \tc{channels} are instances of concurrent types.
The channels left of \lstinline$=>$ are \tc{input polarity} channels.
The channels right of \lstinline$=>$ are \tc{output polarity} channels.

Lines 3 - 9 constitute the \tc{process body}.
We say that the channels listed in line 2 are \tc{in scope} of the process body which means that the process can perform operations, called \tc{process commands}, on these channels.
The process body may be a single process command, as shown in the \lstinline$run$ processes in Examples~\ref{ex:helloworld} and~\ref{ex:serverecho}, or a \lstinline$do$ block of process commands, as shown in \lstinline$broadcast$.
Process commands performed on the same channel can be grouped together with ``\lstinline$on ch do$'' as in Example~\ref{ex:serverecho} on line 3 of \lstinline$client$.
This allows one to omit a repetitive ``\lstinline$on ch$'' after each command.
Certain process commands may only be used as the last process command in a command block, for example an invocation of another process. 
We explain these finer details on process commands in Section~\ref{Sec:concurrent-commands}.

% i don't like this sentence but idk how to make it better.
Different process body definitions can be executed depending on the value of variables.
We will discuss how sequential data can control a process's execution in Section~\ref{sec:seqcontrol}.

Mutually recursive process definitions can be given using a \lstinline$defn$ statement and local process definitions can be given using a \lstinline$where$ statement inside a \lstinline$defn$ statement. 
Details of \lstinline$defn$ and \lstinline$where$ are given in Section~\ref{sec:calling_procs}.

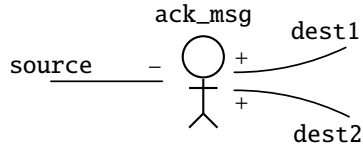
\begin{figure}[h]
\[ \xymatrixcolsep{3pc}
 \xymatrix{ 
 ~  \ar@{-} [r]+<-3ex, +0ex> ^>{-~~}^<{\text{\lstinline$source$}}
 & \overset{\overset{\text{\lstinline$ack_msg$}}{\quad}}{\blackman}
 & ~ \ar@{-} @/^0.25pc/ @<-3ex> [l]+<+3ex, -2.25ex>  _>{~~+}   _<{\text{\lstinline$dest1$}} 
 \ar@{-} @/_0.25pc/ @<+3ex> [l]+<+3ex, +2.25ex>  ^>{~~+}  ^<{\text{\lstinline$dest2$}}
 }  
 \]
 \caption{Diagram of process \lstinline$broadcast$}
 \label{Fig: broadcast}
\end{figure}

We can depict processes using diagrams. 
We will use diagrams to illustrate some of the interactions processes can have along channels.
A diagram of \lstinline$broadcast$ is shown in Figure~\ref{Fig: broadcast}.
We use \lstinline$-$ to denote input polarity and \lstinline$+$ to denote output polarity.
% broadcast

\subsection{Channels}
\label{Sec:Channels}

A process \tc{uses} process commands on a channel to interact with the process plugged into the other end. 
Two processes are connected by (at most) one typed channel that they use with opposite polarities. 
The \tc{type} of a channel defines the interaction that the processes will have over it, and its \tc{polarities} define each process's role in the interaction. 
Thus, the process commands that a process can use on a channel depend on both type and polarity.
The type of a channel can either be explicitly defined in the type signature of a process, or the compiler can infer the type of a channel based on how the processes on either end use it.

\subsubsection{Polarities}
\label{Sec:polarities}

A process is defined with two lists of the channels in its scope: input polarity channels and output polarity channels.
A channel's polarity refers to which \tc{end} of the channel the process uses. 
We say that a process using a channel with output polarity is ``on the left end,'' and a process using a channel with input polarity is ``on the right end.''
The input/output polarity terminology expresses a process-centric perspective, and the left/right end terminology expresses a channel-centric perspective.

Recall Example~\ref{ex:serverecho} in which \lstinline$client$ and \lstinline$server$ are connected by a channel \lstinline$ch$. The \lstinline$client$ uses \lstinline$ch$ with output polarity (the left end) and \lstinline$server$ uses \lstinline$ch$ with input polarity (the right end).
\begin{lstlisting}[label=ex:serverechofull, caption=Server echos Client's message]
proc client = 
	| => ch ->							-- channel ch has output polarity
		on ch do 
			put "Hello Server!" 
			get echo				
			halt

proc server =  
	| ch => ->							-- channel ch has input polarity
		on ch do
			get msg						
			put msg						
			halt
			
 proc run = 
 	| => -> plug						-- program looks like client = ch = server
 		client( | => ch )			-- client is on the left
 		server( | ch => )			-- server is on the right
\end{lstlisting}
Recall that the type of a channel defines the interaction that the processes will have on it.
In the first step of the above interaction, a message travels from left to right.
Recall that the polarity of the channel that each process uses defines that process's role in the interaction.
The \lstinline$client$ uses \lstinline$put$ on its output polarity channel \lstinline$ch$ to send a message. Complementary to \lstinline$put$, \lstinline$server$ uses \lstinline$get$ on its input polarity channel \lstinline$ch$ to receive the message.
In the second step, the message travels in the opposite direction, so the type of \lstinline$ch$ is inferred as the dual of the previous step and the processes use the dual commands.
Finer details on dual channel types and complementary pairs of process commands are given in Section~\ref{subsec:pairs}.

Channel polarities allow one to define unambiguous interactions between processes. 
Without polarities to distinguish their roles, both \lstinline$client$ and \lstinline$server$ could use a \lstinline$get$ command in the first step of the interaction.
Then, they would both wait for a message to be sent by the other, thereby causing a deadlock. 
For example, suppose we do not consider polarity and rewrite our code as follows:
\begin{lstlisting}[label=ex:ambiguityce, caption=Code without channel polarity (this code will not compile)] 
proc client = 
	| ch ->
		on ch do 
			get msg							-- receives server's message  
			put "Hello Server!"	-- sends message to the server      
			halt

proc server =  
	| ch ->
		on ch do
			get msg							-- receives client's message
			put "Hello Client!"	-- sends message to client
			halt

 proc run = 					-- if we could compile this program, it would deadlock!
 	| => -> plug
 		client( | ch )		-- client process has channel ch (no defined polarity)
 		server( | ch )		-- server process has channel ch (no defined polarity)
\end{lstlisting}
Neither process will be able to continue past their \lstinline{get} command because nothing is actually \lstinline{put} on the channel first.
This is precisely what a deadlock is.
In this simple example, it is easy to see the mistake, but in more complicated programs, the type system with polarities has the ability to catch these sorts of errors.
Each built-in channel type in CaMPL specifies the complementary roles of the processes on each end of the channel to ensure it is unambiguous.

\subsubsection{Built-in types}
\label{Sec:channeltypes}

Built-in channel types define the fundamental interactions that are permitted to occur between processes.
The type of a channel can either be explicitly defined in the type signature of a process, or the compiler can infer the type of a channel based on how the processes on either end use it.

As mentioned in Section~\ref{Sec:Processes}, process definitions may include a type signature with three comma-separated lists of sequential types, input concurrent types, and output concurrent types.
Variables and channels are bound to types in the order they appear.

We modify Example~\ref{ex:serverecho} and include the processes' type signatures.
To effectively illustrate the types, we redefine \lstinline$server$ to have a \lstinline$server_id$ variable that it will send back to  \lstinline$client$:

\begin{lstlisting}[label=ex:serverid, caption=Server replies to Client with server id] 
proc client :: | => Put([Char]|Get(Int|TopBot)) = 			-- type signature
	| => ch ->
		on ch do 
			put "Hello Server!"	
			get server_id					
			halt

proc server :: Int | Put([Char]|Get(Int|TopBot)) => = 	-- type signature
	server_id | ch => ->				-- new server_id variable
		on ch do
			get msg							
			put server_id						-- sends server_id back
			halt
			
proc run :: | => = 						-- type signature
 	| => -> plug					
 		client( | => ch )		
 		server( 1234 | ch => )		-- server process created with server_id
\end{lstlisting}
Observe that the type signature of \lstinline$run$ does not contain any types because it is not using any variables or any channels to interact with other processes.
%It creates the channel \lstinline$ch$ in \lstinline$plug$ command to connect the processes.

Notice that \lstinline$ch$ has type \lstinline$Put([Char]|  Get(Int| TopBot))$ which is constructed inductively. 
The outer-most layer \lstinline$Put([Char]| ...)$ defines the first step of the interaction in which a string \lstinline$msg$ travels from left to right (\lstinline$client$ to \lstinline$server$). 
Accordingly, \lstinline$client$ uses \lstinline$put$ on line 4 to send, and \lstinline$server$ uses \lstinline$get$ on line 11 to receive. 
In the next layer \lstinline$Get(Int| ...)$, an integer \lstinline$server_id$ travels from right to left (\lstinline$server$ to \lstinline$client$).
Accordingly, \lstinline$client$ uses \lstinline$get$ on line 5 to receive, and \lstinline$server$ uses \lstinline$put$ on line 12 to send.
Finer details about these commands are given in Section~\ref{subsec:put-get}. 
The inner-most layer \lstinline$TopBot$ is the base case and final step. 
We can \lstinline$close$ the channel or, as above, \lstinline$halt$ by closing the final open channel.
Finer details about these commands are given in Section~\ref{sec:close}. 

\begin{figure}[h]
\[ \xymatrixcolsep{7pc}
 \xymatrix{ 
 \overset{\quad ~ \quad}{\blackman} 
 & \overset{\text{~\lstinline$a:A$~}}{\blackman} \ar[l]_<{+~~}_{\text{\lstinline$ch:Get(A|Ch)$}}_>{~~-} 
 \ar@{~>}[r]^{\text{ \lstinline$get$ / \lstinline$put$} }
 & \overset{\text{~\lstinline$a:A$~}}{\blackman} \ar@{-}[r]^<{~~+}^{\text{\lstinline$ch:Ch$}}^>{-~~} 
 & \overset{\text{~\lstinline$a:A$~}}{\blackman} 
 }  
 \]
 \caption{Message passing on \lstinline$Get$ channels}
 \label{Fig: Get}
\end{figure}
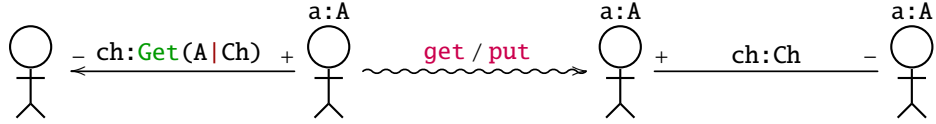

From Example~\ref{ex:serverid}, we can see that a channel of type \lstinline$Put$ indicates messages will travel from left to right, type \lstinline$Get$ indicates messages will travel from right to left, and type \lstinline$TopBot$ indicates the interaction is over and the channel will be closed.
The diagrams in Figures~\ref{Fig: Put} and \ref{Fig: Get} visualize the direction messages travel on each channel type.

%\begin{figure}[h]
%\[  \xymatrix{  
% \blackman  \quad
%  \ar@<8pt>[rrr]^{+ \qquad {\tt Put} \qquad - }
% \ar@<-8pt>@{<-}[rrr]_{+ \qquad {\tt Get} \qquad  -} 
% &  & & \quad \blackman   }  
% \]
% \caption{Direction of messages on \lstinline$Put$ and \lstinline$Get$ types}
% \label{Fig: Put and Get}
%\end{figure}

% PUT
\begin{figure}[h]
\[ \xymatrixcolsep{7pc}
 \xymatrix{ 
 \overset{\text{~\lstinline$a:A$~}}{\blackman} \ar[r]^<{~~+}^{\text{\lstinline$ch:Put(A|Ch)$}}^>{-~~} 
 & \overset{\quad ~ \quad}{\blackman}  
 \ar@{~>}[r]^{\text{ \lstinline$put$ / \lstinline$get$} }
 & \overset{\text{~\lstinline$a:A$~}}{\blackman} \ar@{-}[r]^<{~~+}^{\text{\lstinline$ch:Ch$}}^>{-~~} 
 & \overset{\text{~\lstinline$a:A$~}}{\blackman} 
 }  
 \]
 \caption{Message passing on \lstinline$Put$ channels}
 \label{Fig: Put}
\end{figure}
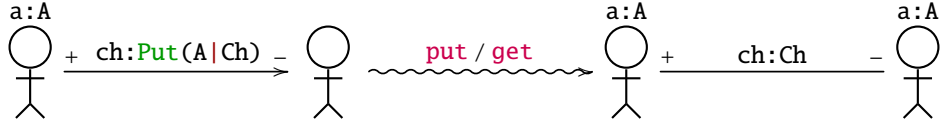

% GET

Next, in Example~\ref{ex:clientfork}, we will consider a compound channel type that allows bundling of two (or more) channels into one.
The types \lstinline$(*)$, called \tc{tensor}, and \lstinline$(+)$, called \tc{par}, come from multiplicative linear logic and allow changes to be made to the network of processes while ensuring no cycles are introduced.
By unbundling the channels, two new channels are created and passed into two new processes.

We again modify Example~\ref{ex:serverecho} to demonstrate \lstinline$(*)$.
We define a new process \lstinline$two_clients$ that uses \lstinline$fork$ on line 9 to unbundle its output channel \lstinline$two_ch$ and create two instances of \lstinline$client$ that both send a message to a modified \lstinline$server$ process. 
This \lstinline$server$ uses \lstinline$split$ on line 15 to unbundle its input channel \lstinline$two_ch$ into two channels, \lstinline$ch1$ and \lstinline$ch2$. After unbundling, \lstinline$two_ch$ is no longer in scope in either process. %The modified example is as follows.

\begin{lstlisting} [label=ex:clientfork, caption=Client forks to create two new processes] 
proc client :: | => Put([Char]|TopBot) = 		
	| => ch ->
		on ch do 
			put "Hello Server!"				  -- client sends string
			halt

proc two_clients :: | => Put([Char]|TopBot) (*) Put([Char]|TopBot) =
	| => two_ch ->
		fork two_ch as								-- unbundles channels ch1 and ch2
			ch1 -> client( | => ch1)		-- creates two new client processes
			ch2 -> client( | => ch2)		-- that each get one channel

proc server :: | Put([Char]|TopBot) (*) Put([Char]|TopBot) => =  
	| two_ch => -> do
		split two_ch into ch1, ch2		-- unbundles channels ch1 and ch2
		on ch1 do											-- interacts with first client
			get msg						
			close
		on ch2 do											-- interacts with second client
			get msg						
			halt
			
proc run :: | => = 
 	| => -> plug						
 		two_clients( | => two_ch )		-- creates a single two_clients process
 		server( | two_ch => )				
\end{lstlisting}
Observe that \lstinline$two_ch$ has type \lstinline$Put([Char]|TopBot)  (*)  Put([Char]|TopBot)$.
The outer-most \lstinline$(*)$ type indicates that the process on the left (\lstinline$two_clients$) will be replaced by two new processes that are both connected to the process on the right (\lstinline$server$).
The next layers indicate how the interactions between the processes on the two new channels will proceed.

What if we wanted the process on the right to be replaced by two new processes instead?
Then, we would use \lstinline$(+)$ which is dual to \lstinline$(*)$.
The diagrams in Figures~\ref{Fig: Tensor} and \ref{Fig: Par} visualize the change in the process network from each channel type.
An example demonstrating a channel with type \lstinline$(+)$ and finer details about the corresponding process commands are given in Section~\ref{sec:split-fork}.

% TENSOR 
\begin{figure}[h]
\[ 
\xymatrixcolsep{7pc}
\xymatrixrowsep{0.1pc}
 \xymatrix{ 
 \overset{\quad ~ \quad}{\blackman} \ar@{-}[r]
^<{~~+}^{\text{\lstinline$ch:Ch1(*) Ch2$}}^>{-~~} 
 & \overset{\qquad}{\blackman}  
 \ar@{~>}[r]^{\text{ \lstinline$fork$ / \lstinline$split$} }
 & \overset{\blackman}{\underset{\blackman}{\quad}} 
 \ar@{-} @/^1pc/ @<+4ex> [r]+<-4ex, -2ex>  ^<{~~+}  ^{\text{\lstinline$ch1:Ch1$}} ^>{-~~}
  \ar@{-} @/_1pc/ @<-4ex> [r]+<-4ex, +2ex>  _<{~~+}   _{\text{\lstinline$ch2:Ch2$}} _>{-~~}
 & \overset{\quad ~ \quad}{\blackman}
 }  
 \]
 \caption{Change in process network from \lstinline$(*)$ channels}
 \label{Fig: Tensor}
\end{figure}

% PAR 
\begin{figure}[h]
\[ 
\xymatrixcolsep{7pc}
\xymatrixrowsep{0.1pc}
 \xymatrix{ 
 \overset{\quad ~ \quad}{\blackman} \ar@{-}[r]
^<{~~+}^{\text{\lstinline$ch:Ch1(+) Ch2$}}^>{-~~} 
 & \overset{\qquad}{\blackman}  
 \ar@{~>}[r]^{\text{ \lstinline$split$ / \lstinline$fork$} }
 & \overset{\quad ~ \quad}{\blackman}
  \ar@{-} @/^1pc/ @<+0.2ex> [r]+<-3ex, +4ex>  ^<{~~+}   ^{\qquad \qquad \text{\lstinline$ch1:Ch1$}} ^>{-~~}
 \ar@{-} @/_1pc/ @<-0.2ex> [r]+<-3ex, -4ex>  _<{~~+}  _{\qquad \qquad \text{\lstinline$ch2:Ch2$}} _>{-~~}
 & \overset{\blackman}{\underset{\blackman}{\quad}} 
 }  
 \]
 \caption{Change in process network from \lstinline$(+)$ channels}
 \label{Fig: Par}
\end{figure}

Lastly, the \lstinline$Neg$ channel type negates a given channel type.
This means that the direction of the interaction is reversed, so it is in the same direction as the \tc{dual} type.
A \lstinline$Neg(Put)$ channel can be used like a \lstinline$Get$ channel (and vice versa), and a \lstinline$Neg((*))$ channel can be used like a \lstinline$(+)$ channel (and vice versa).

Recall process \lstinline$broadcast$ from Example~\ref{ex:broadcastmsg}.
The type signature of \lstinline$broadcast$ is as follows:
\begin{lstlisting}
proc broadcast :: [Char] | Put([Char]|Get([Char]|TopBot)) 										=> Put([Char]|TopBot), Put([Char]|TopBot) = 
	ack_msg | source => dest1, dest2 ->
\end{lstlisting}
Note that the type of \lstinline$dest1$ is \lstinline$Put([Char]|TopBot)$.
We will refer to the destination process on the other end of  \lstinline$dest1$ as \lstinline$proc1$.
Notice that \lstinline$proc1$ must use \lstinline$dest1$ with input polarity.

Suppose we actually want \lstinline$proc1$ to use \lstinline$dest1$ with output polarity.
Instead of changing \lstinline$broadcast$ to connect with \lstinline$proc1$ directly, we can write another process, \lstinline$send_to_dest1$, that connects to \lstinline$proc1$.
Our \lstinline$run$ process would be defined as follows:
\begin{lstlisting}
proc run =
	| => -> plug
		sender( | => source)
		broadcast("message broadcasted" | source => dest1, dest2)
		send_to_dest1( | dest1, neg_dest1 => )	-- type of dest1 is Put 
		proc1( | => neg_dest1)									-- type of neg_dest1 is Neg(Put)
		proc2( | dest2 =>)
\end{lstlisting}	
Messages will travel along \lstinline$dest1$ from left to right and then in the opposite direction on \lstinline$neg_dest1$.
Since \lstinline$dest1$ is type \lstinline$Put$, \lstinline$neg_dest1$ can be \lstinline$Neg(Put)$. 
We define process \lstinline$send_to_dest1$ as follows:
\begin{lstlisting} [label=ex:negdest1, caption=Using a Neg(Put) channel instead of a Get channel] 
proc send_to_dest1 :: | Put([Char] | TopBot), Neg(Put([Char] | TopBot)) => =
	| source, neg_dest1 => -> do	-- send a message on neg_dest1 like a Get type
		on source do								-- receive message from source via broadcast
			get msg
			close
		plug
			neg_dest1, dest1 => -> 		-- create a channel dest1
				neg_dest1 |=| neg dest1 -- negate it and identify with neg_dest1
			=> dest1 -> 							-- use dest1 to forward message to proc1
				on dest1 do	
					put msg
					halt
\end{lstlisting}
To connect to this process, \lstinline$proc1$ would also need to be defined to have a \lstinline$Neg(Put)$ channel instead of a \lstinline$Get$ channel because they are not actually the {\em same} type.
Additionally, \lstinline$proc1$ would need to include code similar to lines 6 - 12 to be able to use the \lstinline$Neg(Put)$ channel.
A complete program that implements \lstinline$proc1$ is given in Appendix~\ref{prog:broadcastmsg}.
Another example demonstrating a \lstinline$Neg$ channel and finer details on \lstinline$neg$ and identification \lstinline$|=|$ are given in Section~\ref{sec:neg}.

%The following are the built-in channel types provided by CaMPL \cite[Section 3.3.2]{Prashanth-thesis}.
Table~\ref{Table:built_in_chs_summary} summarizes the interactions provided by the built-in channel types.
We denote an argument for a sequential type with \lstinline$A$ and a concurrent channel type with \lstinline$Ch$, \lstinline$Ch1$, and \lstinline$Ch2$.
Channel types that take another channel type as an argument are recursive cases. 
The outer-most layer is the first interaction that will take place, and after that, the interaction will continue according to the type of the channel given as the argument.
\begin{table}[!ht]
\begin{center}
\begin{tabular}{| l | l |}
	\hline
	 {\bf Channel type} & {\bf Description of interaction} \\
	 \hline
\lstinline$TopBot$ & End communication on the channel (this is the base case). \\
\lstinline$Put(A|Ch)$ & A message of type \lstinline$A$ travels from left to right (output to input polarity). \\ 
\lstinline$Get(A|Ch)$ & A message of type \lstinline$A$ travels from right to left (input to output polarity). \\
\lstinline$Ch1(*) Ch2$ & Left process becomes two processes both connected to right process. \\
\lstinline$Ch1(+) Ch2$ & Right process becomes two processes both connected to left process. \\
\lstinline$Neg(Ch)$ & Direction of interaction for \lstinline$Ch$ type is reversed. \\ 
		 \hline
\end{tabular}
	\caption{Summary of built-in channel type interactions.}
	\label{Table:built_in_chs_summary}
\end{center}
\end{table}

\subsubsection{Custom types}
\label{Sec:protocols}

We have covered the basic channel types, but the processes we have written so far cannot send variable numbers of messages.
Recall process \lstinline$broadcast$ from Example \ref{ex:broadcastmsg}.
This process was only able to broadcast a single message from the source.
To allow \lstinline$broadcast$ to call itself recursively until all messages are sent, the channel type needs to be able to change from \lstinline$Put$ to \lstinline$TopBot$.
This can be achieved by defining custom channel types, called \tc{protocols} and \tc{coprotocols}, which can be recursive.
Protocols and coprotocols are similar to session types as one process chooses an interaction from a pre-defined set of possible interactions and the other process defines its behaviour for each case.
Non-recursive protocols and coprotocols come from additive linear logic.
Recursive protocols and coprotocols are the concurrent analogue of inductive and coinductive data types in functional programming languages. 
However, a coprotocol is simply a dual protocol, which is not true for sequential inductive and coinductive data types.

Consider the protocol \lstinline$SendMsgs$:
\begin{lstlisting}[label=ex:protocolpassmsgs, caption=Protocol for sending an arbitrary number of messages] 
protocol SendMsgs(A| ) => S =						-- sends messages of type A
	SendMsg :: Put(A|S) => S 							-- handle to send another message
	CloseCh :: TopBot => S 								-- handle to finish sending messages
\end{lstlisting}
This protocol allows an arbitrary number of messages of type \lstinline$A$ to be passed on a \lstinline$Put$ channel.
When we use it as a channel type, we will instantiate the sequential type variable \lstinline$A$.
The \tc{handles} represent the set of valid interactions, or session types, that may take place on the channel. 
The handle \lstinline$SendMsg$ sets the channel type to \lstinline$Put$ to allow another message to be sent.
Once all messages are sent, \lstinline$CloseCh$ can be used to change the channel type to \lstinline$TopBot$.

If we want \lstinline$broadcast$ to also send an acknowledgement message on \lstinline$source$, as it did originally, we need to define another protocol with a handle that sets the channel to send a message and receive a string in response. For example:
 \begin{lstlisting}[label=ex:protocolpassmsgswconf, caption=Protocol for sending messages and receiving acknowledgement messages] 
protocol SendMsgsWithAck(A| ) => S =
	SendMsgWithAck :: Put(A|Get([Char]|S)) => S 	-- handles have new names
	CloseAckCh :: TopBot => S 		
\end{lstlisting}
We need to use different handle names than we did in \lstinline$SendMsgs$ because all handle names must be globally unique.
Neither protocol has concurrent type variables, but a concurrent type variable \lstinline$X$ would be given as \lstinline$SendMsgsWithAck(A| X)$.
The name of a protocol, its handles, and its sequential and concurrent type variables are required to begin with an uppercase alphabet character. 

We can modify \lstinline$broadcast$ to use the \lstinline$SendMsgs$ and \lstinline$SendMsgsWithAck$ protocols as follows:
\begin{lstlisting}[label=ex:recbroadcast, caption=Process that broadcasts messages from a single source]
proc broadcast :: [Char] | SendMsgsWithAck([Char]| ) => SendMsgs([Char]| ), SendMsgs([Char]| ) = 
	ack_msg | source => dest1, dest2 -> do
		hcase source of					-- check whether there is another message
			SendMsgWithAck -> do
				get msg on source		-- receive message
				on dest1 do					-- broadcast message to other processes
					hput SendMsg			-- indicate there is another message
					put msg						-- send that message
				on dest2 do
					hput SendMsg
					put msg
				on source do				-- send acknowledgement to source
					put ack_msg
				broadcast(ack_msg | source => dest1, dest2) -- recurse
			CloseAckCh -> do			-- close all channels and halt
				close source							
				on dest1 do						
					hput CloseCh			-- indicate source is finished
					close							-- close channel
				on dest2 do
					hput CloseCh	
					halt	
\end{lstlisting}
The sequential type variable \lstinline$A$ is instantiated with \lstinline$[Char]$ on line 1 to indicate we are broadcasting strings.
A handle is received on the input polarity channel \lstinline$source$ using \lstinline$hcase$ on line 3.
Process bodies for each handle are defined on lines 4 - 14 and 15 - 22.
On lines 7 and 18, a communication session is initiated on the output polarity channel \lstinline$dest1$ by \tc{activating} it with a handle using \lstinline$hput$.
This sets the channel type as specified by the handle. 
Finer details on these process commands are given in Section~\ref{subsec:hput-hcase}.

The only difference between protocols and coprotocols is the direction that handles travel on the channel.
Protocol handles travel in the same direction as messages on a \lstinline$Put$ channel.
For example, \lstinline$broadcast$ receives handles on \lstinline$source$ and sends handles on its destination channels.
Coprotocol handles travel in the same direction as messages on a \lstinline$Get$ channel.

A \lstinline$send_to_dest1$ process that forwards an arbitrary number of messages to \lstinline$proc1$ could use a \lstinline$Neg(SendMsgs([Char]| ))$ channel.
Alternatively, we could use a coprotocol version of \lstinline$SendMsgs$ instead.
We can define a coprotocol \lstinline$CoSendMsgs$ as follows:
 \begin{lstlisting}[label=ex:coprotocolcopassmsgs, caption=Coprotocol for sending an arbitrary number of messages]
coprotocol S => CoSendMsgs(A| ) =
	CoSendMsg ::  S => Get(A|S)
	CoCloseCh :: S => TopBot
\end{lstlisting}
Notice that the state variable \lstinline$S$ is on the left side of \lstinline$=>$ instead of the right side like in the \lstinline$SendMsgs$ and \lstinline$SendMsgsWithAck$ definitions.
We rewrite \lstinline$send_to_dest1$ using \lstinline$CoSendMsgs$:
\begin{lstlisting} [label=ex:coprotocoldest1, caption=Using a CoSendMsgs channel instead of a Neg(SendMsgs) channel] 
proc send_to_dest1 :: | SendMsgs([Char]| ), CoSendMsgs([Char]| ) => =
	| source, dest1 => -> 
		hcase source of					-- check whether there is another message
			SendMsg -> do
				get msg on source 		
				on dest1 do
					hput CoSendMsg		-- use coprotocol handle
					put msg
				send_to_dest1( | source, dest1 => ) -- recurse
			CloseCh -> do
				close source
				on dest1 do
					hput CoCloseCh
					halt
 \end{lstlisting}
From this example, we can see that a coprotocol streamlines the process of writing code for a negated protocol rather than being a unique feature. 
A complete program based on this example is given in Appendix~\ref{prog:broadcastmsgs}.

Another use case for protocols and coprotocols is the creation of infinitely bundled channel types using \lstinline{(*)} or \lstinline{(+)}.
This is demonstrated in Example~\ref{ex:mutex-protocol} and briefly discussed in Example~\ref{ex:service_handles}.

\subsubsection{Interacting with the outside world using service channels}
\label{Sec:ServiceChannels}

The examples we have considered so far, with the exception of Example~\ref{ex:helloworld}, will not produce any observable effects to a user since they have not been connected to the outside world.
Recall and observe that Example~\ref{ex:helloworld} had the only \lstinline$run$ process that was defined with a non-empty list of channels in its scope.
\tc{Service channels} are the channels that a \lstinline$run$ process can be defined with. The service channel types include \lstinline$Console$, \lstinline$Terminal$, and \lstinline$Timer$. 
They are implemented as special built-in protocols and coprotocols on which processes must always use \lstinline$hput$.
The compiler has built-in service processes which implement the corresponding \lstinline$hcase$ to provide the service.

A \lstinline$Console$ allows a user to interact through the terminal in which the program was run.
A \lstinline$Terminal$ opens a new Alacritty\footnote{https://alacritty.org} terminal for user interaction.
A \lstinline$Timer$ can be used to implement time-out features using controlled non-determinism, which we will discuss in Section~\ref{Sec:races}.
A complete list of the handles for these types is included in Section~\ref{sec:servicehandles}.
For now, we will focus on the following handles:
\begin{lstlisting} [label=ex:servicechs, caption=Service channels, name=services] 
protocol Terminal => S =
    StringTerminalPut :: Put([Char]|S) => S
    StringTerminalGet :: Get([Char]|S) => S 
    StringTerminalClose :: TopBot => S
    
coprotocol S => Console =
    ConsolePut :: S => Get([Char]|S) 
    ConsoleGet :: S => Put([Char]|S) 
    ConsoleClose :: S => TopBot 
\end{lstlisting}

A \lstinline{Prelude} with definitions of the service channel protocols and coprotocols and some simple but useful functions is given in Appendix~\ref{prog:prelude}.
This can be copied to a \lstinline{Prelude.mpl} file that one can \lstinline{include} and use.
Consider a modification of Example~\ref{ex:serverecho} in which we \lstinline{include Prelude} to give \lstinline$client$ a \lstinline$Terminal$ to receive user input and to give \lstinline$server$ a \lstinline$Console$ to log the messages it receives:
\begin{lstlisting} [label=ex:serverclientwithservices, caption=Client and server using service channels]  
include Prelude

proc client :: | => SendMsgs([Char]| ), Terminal =
    | => ch, terminal -> do
        on terminal do								
            hput StringTerminalPut
            put "Hello User! Please enter message in terminal:"
            hput StringTerminalGet
            get msg                   -- receive user input
        on ch do											
            hput SendMsg
            put msg                   -- forward to server
            hput CloseCh
            close
        on terminal do								-- confirm then halt when user is ready
            hput StringTerminalPut
            put "Sent message to server. Press ENTER to close terminal."
            hput StringTerminalGet
            get _											-- wait for user before closing terminal
            hput StringTerminalClose
            halt    
            
proc server :: | SendMsgs([Char]| ), Console => =
    | ch, console => -> do
        hcase ch of
            SendMsg -> do							-- log each message received
                get msg on ch
                on console do
                    hput ConsolePut
                    put "message from user: " ++ msg	-- append msg and print
                server( | ch, console => )
            CloseCh -> do							-- halt after all messages received
                close ch
                on console do
                    hput ConsoleClose
                    halt
                    
proc run :: | Console => Terminal = 
    | console => terminal -> plug
        client( | => ch, terminal )
        server( | ch, console => )
 \end{lstlisting}
Note that on line 30 we use an infix concatenation function from the \lstinline{Prelude}.
This function will be defined in Example~\ref{ex:concat}.
Additionally, a \lstinline{SendMsgs} definition must be added to run this program.
% maybe we add send messages to prelude? is that troll?

\subsection{Non-deterministic processes}
\label{Sec:races}

% priyaa suggested that we use the fork/split code and modify it for the discussion on races.
% we couldn't directly reuse the code
% (in the split fork code, only the client sends a message so that the type signatures are not really really long)
Recall Example~\ref{ex:clientfork} in which two \lstinline$client$ processes each sent a message to \lstinline$server$. 
Notice that \lstinline$server$ received the messages in a pre-determined order.

Consider the case when \lstinline$server$ echos these messages back.
With the features we have used so far, \lstinline$server$ must have the interactions in a pre-determined order.
It would wait for the first \lstinline$client$ before starting its interaction with the second \lstinline$client$, so the second \lstinline$client$ would also wait for the first \lstinline$client$. This is parallel but not concurrent!
%In real world usecases such as ticket-booking system, it is desirable for a server to respond dynamically based on the client first sending a message. 
% we aren't talking about this 
For true concurrency, \lstinline$server$ should dynamically opt to have interactions in the order it receives messages. 
% it's not the client that sends a message first because the server cannot know that. it only knows which one it receives first.
We call this \tc{controlled non-determinism} based on the outcome of a \tc{race} between the channels.
A \lstinline$server$ that calls a locally defined process \lstinline$non_deterministic_server$, which holds a race
using the \lstinline$race$ command, is shown in Example~\ref{ex:nondetserver}:
% defn								-- invoke processes before they are defined using defn
% where defn					-- locally define processes using where
\begin{lstlisting} [label=ex:nondetserver, caption=Server echos each client in the order it receives messages]
defn
	proc server =			-- echos msgs without making clients wait for each other
		| two_ch => -> do
			split two_ch into ch1, ch2
			server_non_deterministic( | ch1, ch2 => )

where defn					-- locally defined processes given after where
	proc server_non_deterministic :: | 																				Put([Char]|Get([Char]|TopBot)), Put([Char]|Get([Char]|TopBot)) => =
		| ch1, ch2 => -> do
			race					-- controlled non-determinism via a race
				ch1 -> server_deterministic( | ch1, ch2 => )		-- client on ch1 wins
				ch2 -> server_deterministic( | ch2, ch1 => )		-- client on ch2 wins
	
	proc server_deterministic :: | 																						Put([Char]|Get([Char]|TopBot)), Put([Char]|Get([Char]|TopBot)) => =
		| winner, loser => -> do
			on winner do	-- interacts with client it received a message from first
				get msg
				put msg
				close
			on loser do		-- interacts with other client
				get msg
				put msg
				halt
\end{lstlisting}
Races can be held for any number of channels on which a process is waiting for a message.
The \lstinline$race$ on line 10 consists of channels \lstinline$ch1$ and \lstinline$ch2$ as indicated by lines 11 and 12, respectively.
A process body is defined for each channel in the race, and the process will execute according to the winning channel's corresponding process body.
If \lstinline$ch1$ wins, line 11 will execute and \lstinline$server_deterministic$ will be called with \lstinline$ch1$ in the argument for the  \lstinline$winner$ channel.
If \lstinline$ch2$ wins, line 12 will execute and \lstinline$server_deterministic$ will be called with \lstinline$ch2$ in the argument for the \lstinline$winner$ channel.

Notice that the type signatures of \lstinline$server_deterministic$ and \lstinline$server_non_deterministic$ are the same. 
This is because races do not change channel types; they only change the order in which the steps of the interactions take place.
Finer details on the \lstinline$race$ command are given in Section~\ref{sec:race-command}.
We discuss locally defined processes in Section~\ref{sec:calling_procs}.
A complete program based on this example is given in Appendix~\ref{prog:non_det_server}.

Another use case for races is a time out feature using the \lstinline{Timer} service channel:
\begin{lstlisting}
coprotocol S => Timer =
    Timer :: S => Get(Int|S (*) Put(()|TopBot))
    TimerClose :: S => TopBot 
\end{lstlisting}
This service is a coprotocol, so we use it on an input polarity channel.
Thus, the \lstinline{Get(Int|...)} type of the \lstinline{Timer} handle allows one to \lstinline{put} an integer with a time limit in microseconds.
To set a timer for 60 seconds, that is 60 000 000.
The \lstinline{(*)} type allows one to split the channel and reuse the recursive part to set arbitrarily many timers.
Finally, the \lstinline{Put(()|...)} type in the non-recursive part can be used in a \lstinline{race}.
%After the specified time limit, the service channel will send an empty tuple \lstinline{()}, so the process will unblock from the race.
We modify Example~\ref{ex:serverclientwithservices} to demonstrate this:
\begin{lstlisting}[label=ex:clienttimeout, caption=Client that times out the Terminal after 60 seconds.]
proc client :: | Timer => SendMsgs([Char]| ), Terminal =
	| timer => ch, terminal -> do
		on terminal do
			hput StringTerminalPut
			put "Hello User! Please enter message. Terminal will time out in 60 s."
			hput StringTerminalGet
		on timer do
			hput Timer
			put 60000000											-- set time limit for 60 seconds
		split timer into new_timer, times_up
		on new_timer do										 	-- no recursion, so close new_timer
			hput TimerClose
			close
		race
			times_up -> do										-- if the timer wins, user timed out
				on times_up do
					get ()
					close
				on ch do												-- don't forward anything to server
					hput CloseCh
					close
				on terminal do									-- halt after msg is finally received
					get msg
					hput StringTerminalPut
					put "User timed out. Message not sent."
					hput StringTerminalClose
					halt
			terminal -> do										-- if user wins, run original code
				get msg on terminal
				on ch do
					hput SendMsg
					put msg
					hput CloseCh
					close
				on terminal do
					hput StringTerminalPut
					put "Sent message to server. Press ENTER to close terminal."
					hput StringTerminalGet
					get _
					hput StringTerminalClose
					close
				on times_up do									-- then close the timer and halt
					get ()
					halt
\end{lstlisting}
The current implementation of the \lstinline{Timer} service does not allow one to cancel a timer.
Thus, even if a user provides input within the time limit, the above process cannot halt until the time limit has elapsed and it receives a unit on line 43.
% yes it would be good to be able to cancel the timer

\subsection{Higher-order processes}
\label{sec:homsgs}

In sequential functional programming, higher-order functions take other functions as input or produce functions as output. 
%The semantics for this requires a type system in which a function can be represented as a type which is based on its type signature.
CaMPL supports a concurrent analogue of this concept. 
\tc{Higher-order processes} 
%Higher-order processes are given by enriching the concurrent part in the sequential part.
% the categorical semantics of higher-order functions is given by cartesian closed categories because the closed part basically gives a self-enrichment, so the concurrent analogue is given by enriching processes in the sequential side.
% idk what happens if you have a process that uses a function that is passed as a message then i guess either the code for that function is packaged up with the process or it's only the name and then hopefully the other process knows how to call that function?? idk
encode other processes as sequential data called \tc{higher-order messages}, pass higher-order messages, and/or decode and invoke an encoded process from a higher-order message.
The semantics for this uses a built-in sequential type \lstinline$Store(S)$ to represent a process based on its type signature \lstinline$S$.

Example \ref{ex:hoprocs} shows higher-order processes passing \lstinline$helloworld$ from Example~\ref{ex:helloworld}. 
A complete program based on this example is given in Appendix~\ref{prog:h.o.hello_world}.
\begin{lstlisting} [label=ex:hoprocs, caption=Passing \lstinline{helloworld} between higher-order processes]
proc ho_sender :: | => Put(Store(|Console=>) | TopBot) =
	| => ch -> do
		on ch do
			put store(helloworld)		-- encode helloworld and send as a message
			halt
			
proc ho_receiver :: | Put(Store(|Console=>) | TopBot), Console => =
	| ch, console => -> do
		on ch do
			get stored_process			-- receive message with encoded helloworld
			close
		on console do
			hput ConsolePut
			put ("Higher order receiver says: Running the stored process")
		use(stored_process)( | console => )		-- decode and invoke helloworld
\end{lstlisting}
Observe that \lstinline$ho_sender$ and \lstinline$ho_receiver$ are connected along a \lstinline$Put(Store(|Console=>) | TopBot)$ type channel.
The \lstinline$Store$ type indicates that a higher-order message is being passed.
The type signature of the encoded process, in this case \lstinline$helloworld$, is indicated within the \lstinline$Store$ type.

To encode a process, the built-in \lstinline$store$ function is used.
This function takes either the name of a previously defined process, as in the above example on line 4, or it takes an anonymous process as an in-line process definition.

To decode and invoke an encoded process, the \lstinline$use$ process command is used as in the above example on line 15.
To \lstinline$use$ an encoded process, a higher-order process must provide instances of variables and channels of the correct types and polarities as specified by the \lstinline$Store$ type signature.

\section{Sequential tier}
\label{Sec:SequentialTypes}

The sequential tier of CaMPL is a basic functional programming language.
Custom sequential types can be defined as data or codata.
The value of variables can be used to conditionally branch the execution of functions and processes.
This section discusses these features, which are related to the sequential type system.
We synthesize content from \cite{Ku18, Po21, Po22}.

\subsection{Functions}
\label{sec:funcs}

Functions are defined using the \lstinline{fun} keyword, and they are called using their name and a comma-separated list of their input arguments enclosed in parentheses.
We define a function \lstinline{isEmpty} that will take a list as input and return a \lstinline$Bool$ depending on whether the input list is empty or not. We will use \lstinline{isEmpty} in Example \ref{ex:serverclientwithservices-branching} to check whether the user has input an empty string.
\begin{lstlisting}[label=ex:isEmpty, caption=Sequential function that checks if a list is empty]
fun isEmpty :: [A] -> Bool =
	[]	-> True
	_		-> False
\end{lstlisting}
  
%infix function syntax (like for example append? concat? as ++ )
We can define infix functions such as concatenation \lstinline{++}, which we used on line 30 of Example~\ref{ex:serverclientwithservices} `\lstinline$put "message from user: "     ++ msg$':
\begin{lstlisting}[label=ex:concat, caption=Infix concatenation function]
fun (++) :: [A],[A] -> [A] =
	a,[]			-> a
	[],a 			-> a
	(b:bs),cs	-> b : (bs ++ cs)
\end{lstlisting}
The parentheses around the function name indicate an infix function is being defined.
The name of an infix function is required to contain only special characters with some constraints \cite{Fo23}.
%https://github.com/campl-ucalgary/campl/blob/main/resources/notes_on_infix_operators.txt

CaMPL also supports mutually recursive function definitions and local function definitions. 
We demonstrate a local function definition in Example~\ref{ex:codata reverse_print}.
This again uses \lstinline$defn$ and \lstinline$where$ in the same way as processes, and we describe the details for processes in Section~\ref{sec:calling_procs}.

\subsection{Data and codata types}
Built-in sequential types include \lstinline$Int$, \lstinline$Bool$, \lstinline$Char$, tuples (e.g., the unit \lstinline$()$ or pairs \lstinline$(Int,Bool)$), lists (e.g., strings \lstinline$[Char]$), and higher-order messages (e.g., \lstinline$Store(Int|Console=>TopBot)$).
We explained and demonstrated the \lstinline$Store$ type in Section~\ref{sec:homsgs}.

%constructor :: data -> create data structure
%case :: data structure -> access data
%record :: data -> build codata structure
%destructor :: codata structure (record?) -> access data
Custom sequential types can be defined as either \lstinline{data} or \lstinline{codata}. 
Data types are defined with constructors that build instances of their data structures.
The data in a data structure can be accessed using \lstinline$case$ statements.
Codata types allow one to represent potentially infinite structures that are evaluated lazily using their \tc{destructors}.
Instances of codata structures are built using \tc{records} that define an implementation of the destructors for that instance.
%Mutually recursive data and codata types can be defined within the same definition using the \lstinline$and$ keyword.

% lists can already grow infinitely, but you can't remove elements from a list infinitely.
% stacks can be defined to allow infinite pops.
Example \ref{ex:codata stack} defines a ``success or failure'' \lstinline{SF} data type (a.k.a. ``Maybe'') and a \lstinline{Stack} codata type which can optionally be implemented to be infinitely popped.
\begin{lstlisting} [label=ex:codata stack, caption=Definition of a Stack codata type using a success/failure data type, numbers=left, name=codata]
data SF(A) -> C =
	SS :: A -> C  							-- Success constructor (a.k.a Just)
	FF :: -> C    							-- Failure constructor (a.k.a Nothing)

codata S -> Stack(A) = 
	Push :: A, S -> S 					-- Push destructor
	Pop :: S -> (SF(A),S)				-- Pop destructor uses SF data type
\end{lstlisting}  
A \lstinline{Stack} stores data of any sequential type \lstinline$A$, and it has two destructors namely \lstinline$Push$ and \lstinline$Pop$. 
New data can be added to the top of the \lstinline{Stack} using \lstinline$Push$.
\lstinline$Pop$ removes the data at the top of the stack and returns a pair of that data, if it exists, and the resulting \lstinline{Stack}.
The \lstinline{SF} data type allows a \lstinline$Pop$ to safely fail if the \lstinline{Stack} is empty.

An instance of a \lstinline{Stack} can be built from a list using \lstinline{(:)} to implement the destructors.
We define the following function \lstinline$listStack$ which returns a record that builds a \lstinline{Stack}:
\begin{lstlisting} [label=ex:record, caption=Function that returns a record to build a Stack from a list, numbers=left]
fun listStack :: [A] -> Stack(A) = 
	cs -> (		-- given a list, returns the record:
		Push := c -> listStack(c:cs),			-- implementation of Push
		Pop := -> case cs of							-- implementation of Pop
			b:bs -> (SS(b), listStack(bs))	-- pop the top of the stack
			[] -> (FF, listStack([]))				-- fail if the stack is empty
		)
\end{lstlisting} 

We use a \lstinline{Stack} in the following \lstinline$reverse_print$ process to collect messages as they are received.
Once all the messages have been received, we will \lstinline$flatten$ and print the \lstinline$stack$ from top to bottom.
Since the most recent message will be on the top of the \lstinline$stack$, the messages will be printed in reverse order.
\begin{lstlisting} [label=ex:codata reverse_print, caption=Printing messages from most recent to least recent]
defn        
	proc reverse_print :: Stack([Char]) | SendMsgs([Char]| ), Console => =
		stack | source, console => -> do
			hcase source of
				SendMsg -> do		-- push messages to stack as they are received
					get msg on source
					reverse_print(Push(msg, stack) | source, console => )
				CloseCh -> do		-- after receiving all messages, print flattened stack
					close source
					on console do
						hput ConsolePut
						put "Printing stack from top to bottom: " ++ flatten(stack)
						hput ConsoleClose
						halt
where
	fun flatten :: Stack([Char]) -> [Char] =
		s -> case Pop(s) of	-- recursively pop strings and concat them
			(SS(str), stack) -> str ++ " " ++ flatten(stack)
			(FF, stack) -> ""
\end{lstlisting}
Note that we don't actually require that \lstinline$stack$ is implemented using the \lstinline$listStack$ record.
To use that implementation, we must initialize \lstinline$stack$ as \lstinline{listStack([])} when \lstinline$reverse_print$ is first called.
A complete program based on this example is given in Appendix~\ref{prog:stack_print}.

Although CaMPL doesn't have built-in syntax for function types, which are required to write higher-order functions, we can define a function codata type, called \lstinline{Func}, with an evaluation destructor, called \lstinline{Eval}.
A record that produces a \lstinline{Func} instance defines the function as the implementation of \lstinline{Eval}.
This allows one to write higher-order functions, such as folds and maps.
We define a function codata type and demonstrate a record that defines a function instance in the following code snippet:
% we have an example number for this, but then we will need to renumber all the examples again, and reupload to github etc.
\begin{lstlisting}
codata F -> Func(A, B) =			-- function codata type
	Eval :: A, F -> B 					-- Eval implementation defines each function

fun sumInts :: -> Func((Int, Int), Int) =	-- produces a Func instance
    -> (Eval := (a, b) -> a + b)					-- using Eval will call the function
\end{lstlisting}

\subsection{Sequential control}
\label{sec:seqcontrol}

Recall that, in Example \ref{ex:serverclientwithservices}, \lstinline$client$  transmitted precisely one message from the user to \lstinline$server$.
To allow a user to input arbitrarily many messages before pressing enter to close the terminal, a stopping condition based on the sequential variable \lstinline{msg} must control the \lstinline$client$'s execution.
Conditional branching of a process's execution uses \lstinline$if$-\lstinline$then$-\lstinline$else$, \lstinline$case$, or \lstinline$switch$ statements or multiple process body definitions that pattern match on the value of variables. 

We modify \lstinline$client$ to prompt the user for input until the user presses enter without typing a message, thereby sending an empty string.
Each time \lstinline$client$ receives a string from the user, it will check if it is empty using the \lstinline$isEmpty$ function defined in Example~\ref{ex:isEmpty}.
\begin{lstlisting} [label=ex:serverclientwithservices-branching, caption=Client execution conditional on user input] 
proc client :: | => SendMsgs([Char]| ), Terminal =
	| => ch, terminal -> do
		on terminal do				-- receive msg from user
			hput StringTerminalPut
			put "Hello User! Enter message in terminal. Press ENTER to close."
			hput StringTerminalGet
			get msg
		if isEmpty(msg)				-- conditional branching on the value of msg
			then do							-- user pressed ENTER, so close channels and halt
				on ch do 
					hput CloseCh
					close 
				on terminal do
					hput StringTerminalClose	
					halt 
			else do							-- user has sent a message, so forward to server
				on ch do											
					hput SendMsg
					put msg
				client( | => ch, terminal)	-- recurse 
\end{lstlisting}   
A complete program based on this example is given in Appendix~\ref{prog:user_input_client_server}.

The above example can also be implemented using \lstinline$case$ on the built-in constructors for \lstinline{Bool}:
\begin{lstlisting}
	case isEmpty(msg) of
		True -> do						-- user pressed ENTER, so close channels and halt
			<lines 10-15>
		False -> do						-- user has sent a message, so forward to server
			<lines 17-20>
\end{lstlisting}

The above example can also be implemented using \lstinline$switch$:\begin{lstlisting}
	switch
		isEmpty(msg) -> do		-- user pressed ENTER, so close channels and halt
			<lines 10-15>
		True -> do						-- user has sent a message, so forward to server
			<lines 17-20>
\end{lstlisting}
Note that \lstinline$switch$ statements streamline the code for an \lstinline$if$-\lstinline$then$-\lstinline$else if$-\lstinline$then$-\lstinline$...$-\lstinline$else$ statement rather than being a unique feature.
That is, they execute the first process body in which the corresponding expression evaluates to \lstinline$True$.
Thus, a catch-all default branch can be defined using \lstinline$True$ as the expression.

The above example can also be implemented by defining another process to mutually recurse with \lstinline{client} and conditionally branch using pattern matching on the instance of \lstinline{msg}: 
\begin{lstlisting}[label=ex:pattern_matching, caption=Conditional branching using pattern matching]
defn
	proc client :: | => SendMsgs([Char]| ), Terminal =
		| => ch, terminal -> do
			on terminal do						-- receive msg from user
				hput StringTerminalPut
				put "Hello User! Enter message in terminal. Press ENTER to close."
				hput StringTerminalGet
				get msg
			pattern_match(msg | => ch, terminal)
	proc pattern_match :: [Char] | => SendMsgs([Char]| ), Terminal =
		"" | => ch, terminal -> do	-- user pressed ENTER. close channels and halt
			on ch do 
				hput CloseCh
				close 
			on terminal do
				hput StringTerminalClose	
				halt 
		msg | => ch, terminal -> do	-- user sent a message, so forward to server
			on ch do											
				hput SendMsg
				put msg
			client( | => ch, terminal)-- recurse 
\end{lstlisting}   
We discuss mutually recursive processes in Section~\ref{sec:calling_procs}.

\section{Process commands manual}
\label{sec:concurrent-operators}
\label{Sec:concurrent-commands}

%\TODO{check that all forward and back references are correct}

Conceptually, we have covered CaMPL's programming features in Sections \ref{Sec:ConcurrentTypes} and \ref{Sec:SequentialTypes}.
Table \ref{Table:summary} summarizes the features and provides references to corresponding examples. 
The finer details of how one can successfully write programs using these features are provided explicitly in this section.
%\comment{the idea is the type system parts can focus on the concepts and be like a nice paper to read and then the proc commands section can be more like a manual because we don't really have a manual manual. they should be able to write some programs after the concurrent and sequential type system sections, and if they have compiler errors or ``why isn't this working'' questions then they can go to the process commands section.}
We provide new examples of processes which coordinate access to a shared memory cell and use a \lstinline{(+)} channel to \lstinline{split} and \lstinline{fork}.
We will also revisit previous examples to emphasize the details of how one uses each process command.

\begin{table}[!ht]
\begin{center}
	\quad \begin{tabular}{ |m{3.3cm} | m{6.75cm} | m{2.4cm} | m{1.4cm} | }
	\hline
		 {\bf Channel type} & {\bf Description} & {\bf Process} \newline {\bf command} & {\bf Example} \\
		 \hline      
		 \lstinline$TopBot$ & Interaction on channel is over.  & \lstinline$close $or \lstinline$halt$ & \ref{ex:serverid} \\
		 \hline      
		 \lstinline$Put(A|Ch)$ & Message of type \lstinline$A$ travels from left to right (output to input polarity).  & \lstinline$put$/\lstinline$get$ & \ref{ex:serverid} \\
		 \hline      
		 \lstinline$Get(A|Ch)$ & Message of type \lstinline$A$ travels from right to left (input to output polarity). & \lstinline$get$/\lstinline$put$ & \ref{ex:serverid} \\
		 \hline      
		 \lstinline$Ch1(*) Ch2$ & Left process becomes two processes both connected to the right process.  & \lstinline$fork$/\lstinline$split$ & \ref{ex:clientfork} \\
		 \hline
		 \lstinline$Ch1(+) Ch2$ & Right process becomes two processes both connected to the left process.  & \lstinline$split$/\lstinline$fork$ & \ref{ex:sendersplit}, \ref{ex:broadcastfork} \\
		 \hline      
		 \lstinline$Neg(Ch)$ & Dual interaction of \lstinline$Ch$ by negating and identifying with another channel.  & \lstinline$neg$ and \lstinline$|=|$  &  \ref{ex:negdest1}, \ref{ex:mem-cell} \\
		 \hline
		 Custom \lstinline$protocol$ & Handles travel from left to right \newline (output to input polarity). & \lstinline$hput$/\lstinline$hcase$ & \ref{ex:protocolpassmsgs}, \ref{ex:recbroadcast} \\
		 \hline
		 Custom \lstinline$coprotocol$ & Handles travel from right to left \newline (input to output polarity). & \lstinline$hcase$/\lstinline$hput$ & \ref{ex:coprotocolcopassmsgs}, \ref{ex:coprotocoldest1} \\
		 \hline
		  N/A & Connect pairs of processes by channels.  & \lstinline$plug$ & \ref{ex:serverecho} \\
		 \hline
		  \lstinline$Put(A|Ch)$ or \newline \lstinline$Get(A|Ch)$ & Next process command block selected with controlled non-determinism.  & \lstinline$race$ & \ref{ex:nondetserver} \\
		 \hline
		  N/A & Encode another process in a sequential \lstinline$Store$ type. & \lstinline$store$ & \ref{ex:hoprocs} \\
		 \hline
		  N/A & Decode and invoke the process \newline encoded in a \lstinline$Store$ type. & \lstinline$use$ & \ref{ex:hoprocs} \\
		 \hline
		  N/A & Conditional branching. & \lstinline$if-then-else$, \lstinline$case$, \lstinline$switch$ & \ref{ex:serverclientwithservices-branching}\\
		 \hline
	\end{tabular}
	\caption{Summary of channel types and process commands.}
	\label{Table:summary}
\end{center}
\end{table}

%\newpage
\subsection{Connecting processes using {\tt plug}}

We have seen the \lstinline$plug$ command used in most of our examples, especially in \lstinline$run$ processes.
This command must be used as the last command in a process command block.

Recall the \lstinline$run$ process from Example \ref{ex:negdest1} that connected \lstinline$broadcast$ to the source and destination processes:
\begin{lstlisting}
proc run =
	| => -> plug						-- plug command is the last (and only) command
		sender( | => source)
		broadcast("message broadcasted" | source => dest1, dest2)
		send_to_dest1( | dest1, neg_dest1 => )
		proc1( | => neg_dest1)				
		proc2( | dest2 =>)		-- proc2 cannot be connected to send_to_dest1
\end{lstlisting}	
The \lstinline$plug$ command, used on line 2 in the above example, connects any number of processes along channels such that the network of connected processes remains acyclic.
That is, the topology of a program is a connected finite acyclic graph consisting of processes as nodes and channels as edges.
Processes which are plugged together run in parallel and may communicate via the channels connecting them. 

Notice that each channel connects exactly two processes.
Channels are linear resources which means that they cannot be arbitrarily created, duplicated, or destroyed.
A channel can only be created in a \lstinline$plug$ command and each end must be plugged into a process.
A channel cannot appear more than twice in a \lstinline$plug$ command either.
Since the \lstinline$plug$ command is the last command in the command block, all existing open channels in scope must be passed to the processes being plugged.
Sequential data is not a linear resource, so it can be duplicated, passed into, and used by any number of processes in a \lstinline$plug$ command.

Two processes can be plugged to each other if and only if they satisfy the following conditions: 1) both process definitions have a channel with the same type and opposite polarities, and 2) plugging the processes will not create a cycle.
In the above example, \lstinline$proc2$ and \lstinline$send_to_dest1$ could not be connected because they both have input polarity channels but no output polarity channels.
Furthermore, changing their process definitions and adding a channel is not possible either because they are both already connected to \lstinline$broadcast$.
Satisfying the second condition also requires that processes are not already connected to each other.

A \lstinline$plug$ command can plug previously defined processes, as above, and/or in-line process definitions.
Recall Example \ref{ex:negdest1}, and notice the in-line process definitions on lines 7 - 8 and 9 - 12:
\begin{lstlisting}
proc send_to_dest1 :: | Put([Char] | TopBot), Neg(Put([Char] | TopBot)) => =
	| source, neg_dest1 => -> do
		on source do						
			get msg
			close
		plug
			neg_dest1, dest1 => -> 		-- channels used by in-line proc definition
				neg_dest1 |=| neg dest1 -- in-line process body
			=> dest1 -> 							-- channels used by in-line proc definition
				on dest1 do							-- in-line process body
					put msg
					halt
\end{lstlisting}
Notice that on lines 7 and 9, the input and output channels used by each process are given, but sequential variables are not given.
Any sequential data in scope at line 6 can be used in the in-line process definitions.

\subsection{Calling processes}
\label{sec:calling_procs}

A process can be invoked by calling it using its name and passing it valid instances of all of the variables and channels in its process definition.
A process must be defined prior to its invocation unless the related processes are contained within a \lstinline{defn} statement.
Recall that we defined mutually recursive processes with \lstinline{defn} in Example~\ref{ex:pattern_matching}.
Processes can also be locally defined using a \lstinline{where} statement within a \lstinline{defn} statement.
Recall that we locally defined processes, which were also within a nested \lstinline{defn} statement, in Example~\ref{ex:nondetserver}.
Locally defined processes can only be called by the processes in the \lstinline{defn} statement above the \lstinline{where} keyword.
For locally defined processes to call each other, as shown in Example~\ref{ex:nondetserver}, the nested \lstinline{defn} statement on line 7 is required.

We have seen many examples with processes called in the \lstinline$plug$ command of a \lstinline$run$ process.
When a process is called outside of a \lstinline$plug$ command, it must be the last command in the command block.
Recall that we called \lstinline$send_to_dest1$ recursively in Example~\ref{ex:coprotocoldest1}:
\begin{lstlisting} 
proc send_to_dest1 :: | SendMsgs([Char]| ), CoSendMsgs([Char]| ) => =
	| source, dest1 => -> 
		hcase source of				
			SendMsg -> do							
				get msg on source 									-- first line of the do block
				on dest1 do
					hput CoSendMsg		
					put msg
				send_to_dest1( | source, dest1 => )	-- process call in the last line
			CloseCh -> do
				close source
				on dest1 do
					hput CoCloseCh
					halt
				
\end{lstlisting}
Notice that when we make the process call on line 9, it is the last command in the \lstinline$do$ block that starts on line 4.
This means that all existing open channels in scope at the time the process is called must be passed to the process.

\subsection{Equating channels using {\tt id} and {\tt neg}}
\label{sec:neg}

The identification command, written as \lstinline$|=|$, equates two channels of the same type and opposite polarities.
This command can be thought of as calling an identity process that ``does nothing'' to a channel.
It must be used as the last command in a command block.

The \lstinline$neg$ command is used to equate two channels of the same polarity if one of them has a \lstinline$Neg$ type.
The \lstinline$neg$ command changes a channel of type \lstinline$X$ to type \lstinline$Neg(X)$ and flips its polarity. 
Thus, after using \lstinline$neg$, both channels will have type \lstinline$Neg(X)$ and opposite polarities, so they can be identified.
The \lstinline$neg$ command can only be used within an identification command.

Negation and identification are particularly useful when channels are passed as resources between multiple processes. 
For example, consider the protocol \lstinline$Passer$ which coordinates exclusive access to a resource that can be instantiated using the concurrent type variable \lstinline$R$:
\begin{lstlisting}[label=ex:mutex-protocol, caption=Passer protocol which uses a Neg channel, numbers=left]
protocol Passer( |R) => S =		-- coordinates access to a channel of type R
	Pass :: R (+) Neg(S) => S 	-- uses a Neg channel to facilitate passing
\end{lstlisting}
We demonstrate \lstinline$Passer$ in Example~\ref{ex:mem-cell} in which two processes take turns accessing a channel \lstinline$mem$ which connects them to a memory cell process called \lstinline$memCell$. 
A complete version of this example in which the memory cell value is set with user input is available in Appendix \ref{prog:mem-cell}.
\begin{lstlisting}[label=ex:mem-cell, caption=Using Passer to pass a channel as a resource, numbers=left]
proc run =
	| => -> plug 
		memCell( "" | mem => )						-- memory cell is initially empty
		memAccess("Ping" | passer => mem)	-- Ping has access first
		memWait("Pong" | => passer)
\end{lstlisting}
The processes are coordinating access to the memory cell \lstinline$memCell$, and they want to write either \lstinline$"Ping"$ or \lstinline$"Pong"$. 
They access the memory cell using channel \lstinline$mem$ which has type \lstinline$MemCh([Char]| )$.
The \lstinline$memCell$ code and \lstinline$MemCh$ protocol are straightforward and their definitions are given in the complete version.
%We will refer to each process using the tag they are writing: for example, \lstinline$"Ping"$ starts with access and \lstinline$"Pong"$ is waiting.

The process with access to \lstinline$memCell$ runs \lstinline$memAccess$.
The process that is waiting runs \lstinline$memWait$.
When access to \lstinline$memCell$ changes, the processes will swap what they are running.
That is, the process running \lstinline$memAccess$ will call \lstinline$memWait$ and vice versa.
To define mutually recursive processes like this, we should use a \lstinline$defn$ statement, as mentioned in Section~\ref{sec:calling_procs}.
However, for clearer explanations, we will consider the process definitions separately.

First, we will consider \lstinline$memAccess$:
\begin{lstlisting}
proc memAccess :: [Char] | Passer( |MemCh([Char]| )) => MemCh([Char]| ) =
	tag | passer => mem -> do		-- note: waiting process has other end of passer
		on mem do									-- access memory cell using mem
			hput MemGet
			get stored_tag
			hput MemPut
			put tag
		hcase passer of Pass -> do	-- block until other process requests to swap
			fork passer as						-- pass memory cell access and swap to waiting
				pass_mem with mem -> 
					pass_mem |=| mem								-- use id to pass mem access
				neg_passer -> plug								-- now wait on other end of passer
					neg_passer, new_passer => ->		
						neg_passer |=| neg new_passer	-- negate other end to id with it	
					memWait(tag | => new_passer)		-- recurse with other end
\end{lstlisting}
We demonstrate how identification can be used to pass a channel as a resource on line 11.
We demonstrate how a channel can be identified with a \lstinline$Neg$ channel using \lstinline$neg$ on line 14.

Next, consider \lstinline$memWait$:
\begin{lstlisting}
proc memWait :: [Char] | => Passer( |MemCh([Char]| )) =
	tag | => passer -> do
		hput Pass on passer										-- request to swap to accessing
		split passer into mem, neg_passer			-- block until other proc passes mem
		plug																	-- need to use other end of passer
			=> neg_passer, new_passer -> 
				neg_passer |=| neg new_passer			-- negate other end to id with it	
			memAccess(tag | new_passer => mem)	-- recurse with other end
\end{lstlisting}
Again, we use  \lstinline$neg$ to identify with a \lstinline$Neg$ channel on line 7.

\subsection{Ending communication using {\tt halt} and {\tt close}}
\label{sec:close}

The \lstinline$close$ and \lstinline$halt$ commands terminate the interaction along a channel and remove it from scope. 
These commands are the only way a channel can be destroyed, and they can only be used on a channel with type \lstinline$TopBot$.
Recall process \lstinline$broadcast$ from Example \ref{ex:broadcastmsg}.
We can reorder the process commands without changing the functionality of the process as follows:

% [label=ex:closehalt, caption=Closing and halting channels in broadcast] 
\begin{lstlisting} 
proc broadcast :: [Char] | Put([Char]|Get([Char]|TopBot)) 										=> Put([Char]|TopBot), Put([Char]|TopBot) = 
	ack_msg | source => dest1, dest2 -> do
		get msg on source						
		put msg on dest1		-- the type of dest1 is TopBot after this line
		close dest1					-- close can be used on a TopBot channel at any point
		put msg on dest2
		close dest2    
		put ack_msg on source	
		halt source					-- halt is used to close the last channel and halt
\end{lstlisting}
Notice that the \lstinline$close$ command cannot be used as the last command in a command block, but it can be used at any other point to close \lstinline$TopBot$ channels.
The \lstinline$halt$ command must be used as the last command in a command block because it halts a process after closing the last \lstinline$TopBot$ channel.
 
\subsection{Complementary pairs }
\label{subsec:pairs}

\tc{Complementary pairs} of process commands are used on opposite polarities of a channel with a certain type.
That is, two processes connected along a channel will each use one command in a complementary pair to perform their role in the interaction.
An extended discussion on the importance of channel polarities is given in Section~\ref{Sec:polarities}.

Each complementary pair has a \tc{blocking} command and a \tc{non-blocking} command.
The process using the blocking command must wait until the other process performs the complementary command to unblock it. 
For example, \lstinline$get$ is a blocking command, so a process is blocked until a message is received from the process performing the complementary \lstinline$put$ which sends the message.
The process using \lstinline$put$ is not blocked, even if the other process hasn't received the message, because message passing is asynchronous.
Table \ref{tab:mpl-constructs} summarizes complementary pairs of commands and whether the process that uses each command is blocked.
 \begin{table}[!h]
\centering
	\begin{tabular}{|l|l|}
	\hline
	{\bf Command} & {\bf Description} \\
	\hline
	\lstinline$put$ & Sends a value on a channel. \\ 
	\lstinline$get$ & Blocks until a value is received on a channel. \\ \hline
	\lstinline$hput$ & Sends a handle on a channel. \\ 
	\lstinline$hcase$ & Blocks until a handle is received on a channel. \\ \hline
	\lstinline$fork$ & Creates two new processes and channels (halts original channel). \\ 
	\lstinline$split$ & Blocks until other ends of new channels have processes (closes original channel). \\ \hline
	\end{tabular}
\caption{Complementary pairs of process commands.}
\label{tab:mpl-constructs}
\end{table}

\subsubsection{Sending and receiving messages using {\tt put} and {\tt get}}
\label{subsec:put-get}

We have used the \lstinline$put$ and \lstinline$get$ process commands to send messages and receive messages, respectively, in most of our examples so far.
Recall Example~\ref{ex:serverid} in which two processes are connected along a channel \lstinline$ch$:
\begin{lstlisting}
proc client =
	| => ch ->									-- uses ch with output polarity
		on ch do 
			put "Hello Server!"			-- client sends string which will unblock server
			get int					
			halt

proc server =
	server_id | ch => ->				-- uses ch with input polarity
		on ch do
			get msg									-- server waits here until msg is received
			put server_id						
			halt
\end{lstlisting}
The \lstinline$get$ command on line 11 blocks \lstinline$server$ until a value is received on channel \lstinline$ch$.
When the value is received, it binds the value to the variable \lstinline$msg$ and proceeds down the process command block. 
Channel \lstinline$ch$ is \tc{used} in the \lstinline$get$ command, and hence, it must be in scope. 
The \lstinline$get$ command cannot be the last command in a command block. 

The \lstinline$put$ command on line 4 sends the value \lstinline$"Hello Server!"$ on channel \lstinline$ch$ and proceeds down the process command block. 
This will allow the \lstinline$server$ process to unblock from its complementary \lstinline$get$ command on line 11.
Again, \lstinline$ch$ is used in the \lstinline$put$ command so it must be in scope, and the \lstinline$put$ command cannot be the last command in a command block. 

These commands can be used on channels with type \lstinline$Put$ or \lstinline$Get$ depending on the direction the message is travelling on the channel.
Recall that messages travel on \lstinline$Put$ channels from left to right (output to input polarity), and messages travel on \lstinline$Get$ channels from right to left (input to output polarity).

If the type of a channel is not explicitly defined, using \lstinline$put$ on the output polarity end and \lstinline$get$ on the input polarity end will force the type of the channel to be \lstinline$Put$.
Dually, using \lstinline$get$ on the output polarity end and \lstinline$put$ on the input polarity end will force the type of the channel to be \lstinline$Get$.
In the above example, the type of channel \lstinline$ch$ is \lstinline$Put$ when the string \lstinline$"Hello Server!"$ is sent and \lstinline$Get$ when the integer \lstinline$server_id$ is sent.
Table \ref{Table: Usage of get and put process commands} describes the usage of \lstinline$put$ and \lstinline$get$ process commands on a channel based on its type and polarity.
\begin{table}[h]
\begin{center}
	(a): \quad \begin{tabular}{  c | c || c }
		  Type & Polarity & Command \\
		 \hline
		 \hline      
		 \lstinline$Put$ & output  & \lstinline$put$ \\
		 \hline
		 \lstinline$Put$ & input  & \lstinline$get$ 
	\end{tabular}
	\quad \quad 
	(b): \quad \begin{tabular}{  c | c || c }
		 Type & Polarity & Command \\
		 \hline
		 \hline      
		 \lstinline$Get$ & output  & \lstinline$get$ \\
		 \hline
		 \lstinline$Get$ & input  & \lstinline$put$ 
	\end{tabular}
	\caption{Usage of \lstinline$put$ and \lstinline$get$ process commands for (a): \lstinline$Put$ and (b): \lstinline$Get$ types.}
	\label{Table: Usage of get and put process commands}
\end{center}
\end{table}

\subsubsection{Channel activation using {\tt hput} and {\tt hcase}}
\label{subsec:hput-hcase}

Custom channel types, called protocols and coprotocols, allow processes to change the type of a channel dynamically at run-time.
Recall the protocol \lstinline$SendMsgs$, coprotocol \lstinline$CoSendMsgs$, and modified \lstinline$send_to_dest1$ process from Examples~\ref{ex:protocolpassmsgs}, \ref{ex:coprotocolcopassmsgs} and \ref{ex:coprotocoldest1}:
\begin{lstlisting}
protocol SendMsgs (A| ) => S =		
    SendMsg :: Put(A|S) => S 				
    CloseCh :: TopBot => S 				

coprotocol S => CoSendMsgs (A| ) =
    CoSendMsg ::  S => Get(A|S)
    CoCloseCh :: S => TopBot

proc send_to_dest1 :: | SendMsgs([Char]| ), CoSendMsgs([Char]| ) => =
	| source, dest1 => -> 
		hcase source of					-- receive protocol handle from broadcast
			SendMsg -> do
				get msg on source 		
				on dest1 do
					hput CoSendMsg		-- send coprotocol handle to dest1
					put msg
				send_to_dest1( | source, dest1 => ) -- recurse
			CloseCh -> do
				close source
				on dest1 do
					hput CoCloseCh
					halt
				
 \end{lstlisting}
The \lstinline$hput$ command on line 15 \tc{activates} channel \lstinline$dest1$ by setting its type according to the handle \lstinline$CoSendMsg$ and proceeds down the process command block.
Since \lstinline$dest1$ is used in the \lstinline$hput$ command, it must be in scope. 
The \lstinline$hput$ command cannot be the last command in a command block. 
Since protocols and coprotocols are concurrent analogues of inductive and coinductive data types, we can think of \lstinline$hput$ as constructing the handle on the channel.
We can also think of the handle as a session type and that the \lstinline$hput$ command initializes a session.

The \lstinline$hcase$ command on line 11 blocks \lstinline$send_to_dest1$ until a handle is received on channel \lstinline$source$.
When the handle is received, \lstinline$send_to_dest1$ proceeds according to the corresponding process command block, i.e. either lines 12 - 17 or lines 18 - 22.
Since \lstinline$source$ is used in the \lstinline$hcase$ command, it must be in scope. 
All variables and channels that are in scope at line 11 are also in scope for each command block.
The \lstinline$hcase$ command must be the last command in a command block. 
This means that all existing open channels in scope, i.e. \lstinline$source$ and \lstinline$dest1$, must be closed or passed into other processes by the end of each block.

The handles used on a channel in a \lstinline$hput$ or \lstinline$hcase$ must be defined with the channel's type. 
Whether a process uses \lstinline$hput$ or \lstinline$hcase$ on a channel depends on its polarity and if the channel type is a protocol or coprotocol.
Recall that handles travel on a protocol channel from left to right (output to input polarity which is the same as \lstinline$Put$), and handles travel on a coprotocol channel from right to left (input to output polarity which is the same as \lstinline$Get$).
Since \lstinline$send_to_dest1$ uses the protocol channel \lstinline$source$ with input polarity, it receives handles using \lstinline$hcase$.
It also uses the coprotocol channel \lstinline$dest1$ with input polarity, so it sends handles using \lstinline$hput$.

If the type of a channel is not explicitly defined, using the handles of an existing protocol in an \lstinline$hput$ command on the output polarity end and an \lstinline$hcase$ command on the input polarity end will force the channel type to be that protocol.
This would not work if the handles belong to an existing coprotocol; in that case, the \lstinline$hput$ and \lstinline$hcase$ commands would need to be used on the opposite ends.

\subsubsection{Multi-process communication using {\tt fork} and {\tt split}}
\label{sec:split-fork}

The \lstinline$fork$ and \lstinline$split$ commands are used to make changes to the network of processes at run-time while ensuring no cycles are introduced.
In Example~\ref{ex:clientfork}, we used these commands to \lstinline$fork$ the process on the left end of a channel into two processes that were both connected to the process on the right.
We will consider another example using these commands in which the process on the left uses \lstinline$split$ to talk to two new processes on the right instead.

Recall process \lstinline$broadcast$ from Example~\ref{ex:broadcastmsg}.
We will consider the corresponding \lstinline$sender$ process that is connected along the \lstinline$source$ channel:
\begin{lstlisting}[label=ex:sender, caption=Process that sends messages to broadcast] 
proc sender =
	| => source ->
		on source do
			put "Hi everyone!"
			get ack_msg
			halt
\end{lstlisting} 

Suppose we want \lstinline$sender$ to first send a message that both destination processes will receive and then split the channel and send different messages to each process.
We will omit acknowledgement messages in this modified example for simplicity.
\begin{lstlisting} [label=ex:sendersplit, caption=Splitting the source channel, name=splitforksource] 
proc sender =
	| => source -> do
		on source do
			put "Hi everyone!"										-- msg sent to both procs
		split source into source1, source2			-- unbundle channels
		on source1 do														-- first send msg to 1
			put "Hello proc1"
			close
		on source2 do														-- then send msg to 2 and halt
			put "Hello proc2"
			halt
\end{lstlisting} 
The \lstinline$split$ command on line 5 removes the channel \lstinline$source$ from scope, creates two new channels \lstinline$source1$ and \lstinline$source2$ in scope, and blocks \lstinline$sender$ until two new processes are created and connected on the new channels.
After the \lstinline$split$ command unblocks, \lstinline$sender$ will proceed down the process command block, which means the process commands used on the two new channels in lines 6 - 8 and 9 - 11 will be executed sequentially.
The \lstinline$source$ is used in the \lstinline$split$ command, so it must be in scope at line 5, but the command removes it from scope.
The two new channels will have the same polarity as the original channel.
The \lstinline$split$ command cannot be the last command in a command block. 

The corresponding \lstinline$broadcast$ process will need to \lstinline$fork$ its \lstinline$source$ channel and create two new processes to concurrently broadcast the second round of messages:
%We can define this modified \lstinline$broadcast$ process as follows:
\begin{lstlisting} [label=ex:broadcastfork, caption=Forking the source channel, name=splitforksource] 
proc broadcast = 
	ack_msg | source => dest1, dest2 -> do
		get msg1 on source				
		put msg1 on dest1									-- broadcast msg1 to both processes
		put msg1 on dest2
		fork source as										-- unbundle channels
			source1 with dest1 -> do				-- send msg2a to 1
				get msg2a on source1
				on dest1 do
					put msg2a
					close
				halt source1
			source2 with dest2 -> do				-- send msg2b to 2
				get msg2b on source2
				on dest2 do
					put msg2b
					close
				halt source2
\end{lstlisting} 
The \lstinline$fork$ command on line 17 removes the channel \lstinline$source$ from scope and creates two new channels that are each passed into one of the two process bodies on lines 18 - 23 and 24 - 29.
After the \lstinline$fork$ command, the two process bodies each have one of the two new channels in scope and process commands in each process body will be executed concurrently.
Channel \lstinline$source$ is used in the \lstinline$fork$ command, so it must be in scope at line 17, but it is removed from scope by the command.
The two new channels will have the same polarity as the original channel.

The \lstinline$fork$ command must be the last command in a command block. 
This means that all existing open channels in scope at line 17, i.e. \lstinline$dest1$ and \lstinline$dest2$, must be passed into the process bodies.
On line 18, \lstinline$dest1$ is passed into the the first process body, and on line 24, \lstinline$dest2$ is passed into the second process body.
All variables that are in scope when the \lstinline$fork$ command is made are also in scope for each process body.
For example, if this process was sending acknowledgement messages, the variable \lstinline$ack_msg$ could be used in both process bodies.

These commands can be used on channels with type \lstinline$(*)$, ``tensor,'' or \lstinline$(+)$, ``par,'' depending on which end of the channel has one process \lstinline$fork$ into two processes.
Recall that a \lstinline$(*)$ channel means the process on the left forks (using the channel with output polarity), and a \lstinline$(+)$ channel means the process on the right forks (using the channel with input polarity).

Again, if the type of a channel is not explicitly defined, using \lstinline$fork$ on the output polarity end and \lstinline$split$ on the input polarity end will force the type of the channel to be \lstinline$(*)$.
Dually, using \lstinline$split$ on the output polarity end and \lstinline$fork$ on the input polarity end will force the type of the channel to be \lstinline$(+)$.
In Example \ref{ex:sendersplit}, the type of \lstinline$source$ is \lstinline$(+)$ because \lstinline$sender$ is on the output polarity end and \lstinline$broadcast$ is on the input polarity end.
Table \ref{Table: Usage of fork and split process commands} describes the usage of the \lstinline$fork$ and \lstinline$split$ commands on a channel based on its type and polarity.
\begin{table}[h]
\begin{center}
	(a): \quad \begin{tabular}{  c | c || c }
		 Type & Polarity & Command \\
		 \hline
		 \hline      
		 \lstinline$(*)$ & output  & \lstinline$fork$ \\
		 \hline
		 \lstinline$(*)$ & input  & \lstinline$split$ 
	\end{tabular}
	\quad \quad 
	(b): \quad \begin{tabular}{  c | c || c }
		  Type & Polarity & Command \\
		 \hline
		 \hline      
		 \lstinline$(+)$ & output  & \lstinline$split$ \\
		 \hline
		 \lstinline$(+)$ & input  & \lstinline$fork$ 
	\end{tabular}
	\caption{Usage of \lstinline$fork$ and \lstinline$split$ process commands for (a): \lstinline$(*)$ and (b): \lstinline$(+)$ types.}
	\label{Table: Usage of fork and split process commands}
\end{center}
\end{table}

\subsection{Controlled non-determinism using {\tt race}}
\label{sec:race-command}

The \lstinline$race$ command can be used to race channels on which a blocking command (see Section \ref{subsec:pairs}) is the next command that will be used on each channel.
For example, if a \lstinline$get$ command is the next command that will be used on input polarity \lstinline$Put$ channels or output polarity \lstinline$Get$ channels, those channels can be used in a race just before any \lstinline$get$ commands are used.
We recently added support for racing on \lstinline$hcase$ and \lstinline$split$ commands. 
Previous versions only supported \lstinline$get$ commands, so ensure the most recent version of the compiler is being used for full functionality.

A race allows the process to unblock when any channel in the race unblocks, e.g. when the first message is received on any of the channels in the race.
We call this channel the \tc{winner} of the race.
When the process unblocks, it will execute the winner's corresponding process body.
The rest of the process bodies are ignored.
Since only one of the race's process bodies will be executed, all existing open channels in scope must be closed or passed into another process by the end of each process body.
The \lstinline$race$ command must be the last command in a command block.
We demonstrate the \lstinline$race$ command in the examples in Section~\ref{Sec:races}.

\subsection{Service channel handles}
\label{sec:servicehandles}

%Services are built-in types, so they do not need to be defined in a CaMPL program file to be used.
%However, previous versions of the CaMPL compiler did not contain service channel definitions and required them to be defined in a CaMPL file.
%As such, CaMPL programs with user-defined services will still compile, but a warning will be printed indicating that the user-defined service has been overwritten by the built-in definition.
A current list of service channel handles supported by CaMPL is given in Example~\ref{ex:service_handles}.
%Note that one can define service channels with any subset of these handles, restricted only by whether they are protocol or coprotocol handles.
%However, since handle names must be globally unique, each handle can only be defined for one service channel type at a time.
%One can use any name for a service channel type definition.
%For example, we have used the name \lstinline{Terminal} to group all handles related to Alacritty terminals.
\begin{lstlisting}[label=ex:service_handles, caption=Complete list of service channel handles]
protocol Terminal => S =
    StringTerminalPut :: Put([Char]|S) => S
    StringTerminalGet :: Get([Char]|S) => S 
    StringTerminalClose :: TopBot => S
    
    IntTerminalPut :: Put(Int|S) => S
    IntTerminalGet :: Get(Int|S) => S 
    IntTerminalClose :: TopBot => S

    CharTerminalPut :: Put(Char|S) => S
    CharTerminalGet :: Get(Char|S) => S 
    CharTerminalClose :: TopBot => S

coprotocol S => Console =
    ConsolePut :: S => Get([Char]|S) 
    ConsoleGet :: S => Put([Char]|S) 
    ConsoleClose :: S => TopBot 
    ConsoleStringTerminal :: S => S (*) Neg(Terminal)

    IntConsolePut :: S => Get(Int|S) 
    IntConsoleGet :: S => Put(Int|S) 
    IntConsoleClose :: S => TopBot 

    CharConsolePut :: S => Get(Char|S) 
    CharConsoleGet :: S => Put(Char|S) 
    CharConsoleClose :: S => TopBot 

coprotocol S => Timer =
    Timer :: S => Get(Int|S (*) Put(()|TopBot))
    TimerClose :: S => TopBot 
\end{lstlisting} 
Notice that the handle \lstinline{ConsoleStringTerminal} on line 18 allows one to \lstinline{split} a \lstinline{Console} channel to generate arbitrarily many \lstinline{Terminal} channels.
We have included a Prelude file in Appendix~\ref{prog:prelude} that provides service channel definitions.
A modification to the compiler that will remove the necessity of user-defined services is currently in progress.

\section{Mathematical underpinning}
\label{Sec:math_underp}
% convey that these constructs are coming from a mathematical foundation and not just ``out of thin air''

CaMPL implements a type system based on linear logic that was described in Cockett and Pastro’s paper ``The Logic of Message Passing'' \cite{CoP07}. The primary result of their paper was defining and proving a concurrent analogue of the functional programming Curry-Howard-Lambek correspondence or proofs-as-programs principle \cite{Howard1980, Lambek1969, Lambek1972}. They designed a two-tiered logic consisting of a sequential tier, called the message logic, that interacts with a concurrent tier, called the message-passing logic. 
The message logic is a simple logic that corresponds to sequential functional programming.
The message-passing logic corresponds to programming concurrent processes which use asynchronous message passing along channels. They provided the term calculus for both logics \cite[Section 2.1, Section 3.1]{CoP07} which are implemented by CaMPL. We will discuss the syntax of CaMPL in reference to the two-tiered logic after briefly describing the categorical semantics of message passing.

The categorical semantics of message passing is given by a {\em linear actegory} in which a category of messages and sequential functions acts on a category of channels and concurrent processes in two directions.
The covariant action passes messages in the forward direction, which we described in this article as ``left to right,'' and the contravariant action passes messages in the backward direction, which we described as ``right to left.''

\subsection{Categorical semantics}

% note that in the logic of message passing, at the beginning of section 5: categorical semantics, on page 33, it says that the semantics are a ``linear additive actegory'' which it defines as ``We shall say that a linear A-actegory is A-additive in case the monoidal category A is a distributive monoidal category (i.e., it has coproducts over which the tensor distributes) and the covariant action preserves these coproducts while the contravariant action turns them into products.'' HOWEVER robin has also told me that i am wrong when i talk about the additive part, so i have no idea. i think the distributive part is fine? just not the additive part??}

A linear $\A$-actegory minimally consists of a monoidal category $\A$ acting covariantly and contravariantly on a linearly distributive category 
%(LDC) 
$\X$, satisfying certain coherences~\cite{CoP07, CS97}. 
The categorical semantics of CaMPL is modelled by a symmetric linear $\A$-actegory in which a distributive symmetric monoidal category $\A$, which defines the semantics of the sequential tier, acts on a symmetric linearly distributive category $\X$, which defines the semantics of the concurrent tier. Linearly distributive categories are the categorical semantics of multiplicative linear logic \cite{CS97, Sri21}. 

In the monoidal category $\A$, objects correspond to sequential types, or message types, and morphisms correspond to sequential functions.
Since $\A$ is a distributive monoidal category, it has coproducts over which its tensor distributes.
The tensor in $\A$ corresponds to tuples and the coproducts correspond to data types with multiple constructors.

In the linearly distributive category $\X$, objects correspond to concurrent types, or channel types, and morphisms correspond to concurrent processes.
Since $\X$ is a linearly distributive category, it has two functors $\ox$ and $\oa$, corresponding to the multiplicative conjunction $\ox$ and the multiplicative disjunction $\oplus$ of linear logic:
\[ \otimes: \X \times \X \to \X \quad \quad \text{ and } \quad \quad \oplus: \X \times \X \to \X\]
These functors define the bundled channel types.
The \(\otimes\) functor corresponds to the \lstinline{(*)} channel type, and the \(\oplus\) functor corresponds to the  \lstinline{(+)} channel type.

The interaction of $\A$ and $\X$ is defined by two {\em action} functors:
\[ \circ: \A \times \X \to \X \quad \quad \text{ and } \quad \quad \bullet: \A^{\sf op} \times \X \to \X\]
These functors describe how messages are passed on channels.
The \(\circ\) action corresponds to the \lstinline{Put} channel type, and the \(\bullet\) action corresponds to the the \lstinline{Get} channel type.
The $\circ$ functor is the left parametrized left adjoint of $\bullet$ in the sense that the following is a parametrized adjunction for all message types \(A\) in \(\A\):
\[ \quad A \circ - \dashv A \bullet - : \X \to \X \]
The adjunction signifies that a process sending or receiving a message can do so from either end of a channel. 

The categorical semantics of CaMPL differ slightly from what was described in Cockett and Pastro’s paper.
Since \lstinline{TopBot} is the unit channel type, we only have one unit instead of two, so $\X$ is actually an isomix category \cite{CS97a}.
Furthermore, the \lstinline{Neg} channel type gives each channel type a linear adjoint which means that  $\X$ is a $\ast$-autonomous category \cite{barr}.
The categorical semantics of non-recursive protocols are coproducts, and coprotocols are products \cite{CoS01, CoP04}.
For recursive protocols (and coprotocols), the categorical semantics is given by pairs of initial functor algebras and final functor coalgebras \cite{Ye12}.
The categorical semantics for non-deterministic processes is obtained by enriching the concurrent semantics, $\X$, in sup-lattices by taking the power set of each hom-set of morphisms \cite{xanna}.
This allows a process to be represented by a set of processes that define its possible executions.
The categorical semantics for higher-order processes is obtained by enriching the concurrent semantics, $\X$, in the sequential semantics, $\A$ \cite{No25}.
This allows a process to be represented by a sequential type so that it can be passed as a (higher-order) message.

\subsection{Two-tiered logic}

The two-tiered logic gives a type system for concurrent processes that use message passing as their concurrency primitive \cite{CoP07}. 
Its inference rules govern operations within each tier and interactions between the two tiers. 
%The two-tiers include the logic of messages and logic of message-passing.
The two tiers provide a clean separation of computation and communication operations, so the complexities associated with each can be addressed effectively.

\subsubsection{The logic of messages}

The logic of messages, {\bf Msg}, is the sequential tier, and the proofs in this logic correspond to sequential programs. 
It is concerned with the generation of messages and represents the logic of computation. 
The logic is presented in a Gentzen-style sequent calculus. A sequent takes the form
\[ \Phi \vdash A \]
where the antecedent (or the {\em context}), $\Phi$, is a comma-separated list of formulas and the succedent $A$ is a single formula.
The inference rules for {\bf Msg} are given in Figure \ref{Fig:MsgRules}. In the inference rules, ``subs'' stands for substitution and is the sequential cut rule which corresponds to function composition. 

%\begin{figure}[h]
%	\centering
%	\includegraphics[scale=0.5]{figs/MsgRules.PNG}
%	\caption{Inference rules for {\bf Msg}}
%	\label{Fig:MsgRules}
%\end{figure}
%% figure 1 from the logic of message passing

\begin{figure}[h]
	\centering
	\mprset{leftstyle={\footnotesize\normalfont}}
	{\scriptsize	 
	% sets the style of the labels of the inference rules
	% WHY DOES IT SAY DEF?????
	\begin{align*}
		& \inferrule*[Left=\text{axiom}]{~}{\Phi \vdash A}
		&& \inferrule*[Left=\text{subs}]{\Phi \vdash A \and \Psi_1, A, \Psi_2 \vdash B}{\Psi_1, \Phi, \Psi_2 \vdash B}\\
		& \inferrule*[Left=\(\ast_l\)]{\Phi, A, B \vdash C}{\Phi, A \ast B \vdash C}
		&& \inferrule*[Left=\(\ast_r\)]{\Phi \vdash A \and \Psi \vdash B }{\Phi, \Psi \vdash A \ast B}\\
		& \inferrule*[Left=\(I_l\)]{\Phi \vdash A}{\Phi, I \vdash A} 
		&& \inferrule*[Left=\(I_r\)]{~}{~ \vdash I} \\
		& \inferrule*[Left=\(\text{coprod}\)]{\Phi, A \vdash C \and \Phi, B \vdash C }{\Phi, A + B \vdash C }
		&& \inferrule*[Left=\(\text{inj}_l\)]{\Phi \vdash A}{\Phi \vdash A + B} \\
		& \inferrule*[Left=\(0\)]{~}{\Phi, 0 \vdash A}
		&& \inferrule*[Left=\(\text{inj}_r\)]{\Phi \vdash B}{\Phi \vdash A + B} 
	\end{align*}
	}
	\caption{Inference rules for {\bf Msg}.}
	\label{Fig:MsgRules}
\end{figure}

\subsubsection{The logic of message passing}
	
The logic of message passing, {\bf PMsg}, is the concurrent tier, and the proofs in this logic correspond to concurrent programs. 
It is concerned with interactions between processes over channels and represents the logic of communication. 
A sequent of {\bf PMsg} corresponds to a process, so it has three components all of which are unordered lists: $\Phi$ is the sequential context of message types, $\Gamma$ is input polarity channels types, and $\Delta$ is output polarity channel types. A sequent in {\bf PMsg} is of the following form:
\[ \Phi \mid \Gamma \Vdash \Delta \]	
The inference rules for {\bf PMsg} are shown in Figure \ref{Fig:PMsgRules}. 
Each rule corresponds to a process command.
For example, the cut rule corresponds to the \lstinline{plug} command.
Complementary pairs of process commands arise from the two-sidedness of the inference rules.

The top half of the rules come from multiplicative linear logic and define the valid changes one can make to the process network.
For example, we will consider the rules corresponding to \lstinline{split} and \lstinline{fork}. 
Recall that $\otimes$ and $\oplus$ correspond to the \lstinline{(*)} and \lstinline{(+)} types, respectively. 
The rule $\otimes_l$ is the same as using \lstinline{split} on an input channel, and the rule $\otimes_r$ is the same as using \lstinline{fork} on an output channel. 
The rule $\oplus_l$ is the same as using \lstinline{fork} on an input channel, and the rule $\oplus_r$ is the same as using \lstinline{split} on an output channel.

The bottom half of the rules include the action rules, which pertain to message passing, sequential context rules, and sequential control rules.
For example, we will consider the rules corresponding to \lstinline{get} and \lstinline{put}.
Recall that $\circ$ and $\bullet$ correspond to the \lstinline{Put} and \lstinline{Get} types, respectively. 
The rule $\circ_l$ is the same as using \lstinline{get} on an input channel, and the rule $\circ_r$ is the same as using \lstinline{put} on an output channel.
The rule $\bullet_l$ is the same as using \lstinline{put} on an input channel, and the rule $\bullet_r$ is the same as using \lstinline{get} on an output channel.

Protocols, coprotocols, races, and higher-order message passing were not discussed in Cockett and Pastro's paper.
The inference rules for non-recursive protocols and coprotocols come from additive linear logic, which are similar to the additive rules in \cite{CoP04}.
Recursive instances of protocols and coprotocols use proof boxes for fixed point combinators \cite{Ye12}.
Races do not change the type of a process, so an inference rule for a race only requires specifying the different possible executions as different orderings of inference rule applications depending on the outcome of the race.
Higher-order message passing has two additional rules for \lstinline{store} and \lstinline{use} which are given in \cite{No25}.

\begin{figure}[h]
	\centering 
	{\scriptsize	 
	\mprset{leftstyle={\footnotesize\normalfont}}
	\begin{align*}
	\inferrule*[Left=\text{cut}]{\Phi \mid \Gamma_1 \Vdash \Delta_1, X \and \Psi \mid X, \Gamma_2 \Vdash \Delta_2}{\Phi, \Psi \mid \Gamma_1, \Gamma_2 \Vdash \Delta_1, \Delta_2}
	\end{align*}
	\vspace{-3em} 
	\begin{align*}
		& \inferrule*[Left=\text{atom id}]{~}{\emptyset \mid X \Vdash X} 
		&& \inferrule*[Left=\text{axiom}]{~}{\Phi \mid \Gamma \Vdash \Delta}\\
		& \inferrule*[Left=\(\otimes_l\)]{\Phi \mid \Gamma, X, Y \Vdash \Delta}{\Phi \mid \Gamma, X \otimes Y \Vdash \Delta} 
		&& \inferrule*[Left=\(\otimes_r\)]{\Phi \mid \Gamma_1 \Vdash \Delta_1, X \and \Psi \mid \Gamma_2 \Vdash Y, \Delta_2}{\Phi, \Psi \mid \Gamma_1, \Gamma_2 \Vdash \Delta_1, X \otimes Y, \Delta_2} \\
		& \inferrule*[Left=\(\oplus_l\)]{\Phi \mid \Gamma_1, X \Vdash \Delta_1 \and \Psi \mid Y, \Gamma_2 \Vdash \Delta_2}{\Phi, \Psi \mid \Gamma_1, X \oplus Y, \Gamma_2 \Vdash \Delta_1, \Delta_2} 
		&& \inferrule*[Left=\(\oplus_r\)]{\Phi \mid \Gamma \Vdash X, Y, \Delta}{\Phi \mid \Gamma \Vdash X \oplus Y, \Delta} \\
		&  \inferrule*[Left=\(\top_l\)]{\Phi \mid \Gamma \Vdash \Delta}{\Phi \mid \Gamma, \top \Vdash \Delta} 
		&& \inferrule*[Left=\(\top_r\)]{~}{\emptyset \mid ~ \Vdash \top}\\
		& \inferrule*[Left=\(\bot_l\)]{~}{\emptyset \mid \bot \Vdash } 
		&& \inferrule*[Left=\(\bot_r\)]{\Phi \mid \Gamma \Vdash \Delta}{\Phi \mid \Gamma \Vdash \bot, \Delta}\\ \\
		& \inferrule*[Left=\(\circ_l\)]{\Phi, A \mid \Gamma, X \Vdash \Delta}{\Phi \mid \Gamma, A \circ X \Vdash \Delta} 
		&& \inferrule*[Left=\(\circ_r\)]{\Phi \vdash A \and \Psi \mid \Gamma \Vdash X, \Delta}{\Phi, \Psi \mid \Gamma \Vdash A \circ X, \Delta} \\
		& \inferrule*[Left=\(\bullet_l\)]{\Phi \vdash A \and \Psi \mid \Gamma, X \Vdash \Delta}{\Phi, \Psi \mid \Gamma, A \bullet X \Vdash \Delta}
		&& \inferrule*[Left=\(\bullet_r\)]{\Phi, A \mid \Gamma \Vdash X, \Delta}{\Phi \mid \Gamma \Vdash A \bullet X, \Delta} \\
		& \inferrule*[Left=\(\ast\)]{\Phi, A, B \mid \Gamma \Vdash \Delta}{\Phi, A \ast B \mid \Gamma \Vdash \Delta}
		&& \inferrule*[Left=\(I\)]{\Phi \mid \Gamma \Vdash \Delta}{\Phi, I \mid \Gamma \Vdash \Delta} \\
		& \inferrule*[Left=\(\text{coprod}\)]{\Phi, A \mid \Gamma \Vdash \Delta \and \Phi, B \mid \Gamma \Vdash \Delta}{\Phi, A + B \mid \Gamma \Vdash \Delta}
		&& \inferrule*[Left=\(0\)]{~}{\Phi, 0 \mid \Gamma \Vdash \Delta}
	\end{align*}
	\vspace{-2.5em}
	\begin{align*}
		\inferrule*[Left=\text{subs}]{\Phi \vdash A \and \Psi, A \mid \Gamma \Vdash \Delta}{\Phi, \Psi \mid \Gamma \Vdash \Delta}
	\end{align*}
	}
	\caption{Inference rules for {\bf PMsg}.}
	\label{Fig:PMsgRules}
\end{figure}

%% SCREENSHOT FROM THE LOGIC OF MESSAGE PASSING:
%\begin{figure}[p]
%	\centering
%	\includegraphics[scale=0.6]{figs/PMsgRules.PNG}
%	\caption{Inference rules for {\bf PMsg} \label{Fig:PMsgRules}}
%\end{figure}
%% figure 4

%\begin{figure}[p]
%	\centering
%	\includegraphics[scale=0.6]{figs/programming-syntax.png}
%	\caption{Programming syntax for {\bf PMsg} \label{Fig:programming-syntax}}
%\end{figure}
%% figure 6

\section{Conclusion}

We hope that a reader can now enthusiastically answer the questions  ``what was the type of channel \lstinline$ch$ from Example~\ref{ex:serverecho}?'' and ``what is channel polarity?''
Additionally, we hope that a reader was engaged by the examples we used to showcase the programming features of CaMPL.
To maximize enjoyment of the examples, we suggest compiling and running them!
The current implementation of CaMPL is available at \url{https://campl-ucalgary.github.io/}.
A reasonably current version is available on the online compiler  \url{https://campl-app.vercel.app/}.
Complete programs that can be directly copied and pasted are given in Appendix~\ref{sec:appendix}.

The CaMPL project is still very much in progress.
As such, CaMPL is a proof-of-concept langauge for its categorical semantics.
On the implementation side, we are working on updating the compiler to be more user friendly.
We also want to add features to work with processes that are distributed over multiple devices and connected by a network. 
Finally, of particular interest and novelty is adding type classes to the concurrent type system of CaMPL. Recall that, in Haskell, the type class system increases the expressiveness of the language significantly by enabling adhoc polymorphism. 

On the categorical semantics side, we are working on a precise semantics for non-determinism that considers how races interact with the other features.
We are also considering a categorical semantics for message passing between quantum processes \cite{QMP23}.
This will contribute to the development of programming languages for distributed quantum computing over a quantum internet.

\printbibliography

\begin{appendix}
\section{Examples: Complete programs}
\label{sec:appendix}

The current implementation of CaMPL is available at \url{https://campl-ucalgary.github.io/} -- our website has detailed instructions on how to run CaMPL code.
A reasonably current version is available on the online compiler  \url{https://campl-app.vercel.app/}.
These example programs, along with many others, are also available on our Github:
\url{https://github.com/campl-ucalgary/campl/tree/main/MPLCLI/examples/complete-appendix-programs}

\subsection{Prelude}
\label{prog:prelude}

CaMPL has a module system that allows one to include and use code defined in other files \cite{Fo23mod}.
This \lstinline{Prelude.mpl} file combines code from Examples \ref{ex:servicechs}, \ref{ex:isEmpty}, \ref{ex:concat}, and \ref{ex:service_handles}.
We encourage readers to write additional functions in their version of the Prelude as an exercise in CaMPL programming.
We demonstrate how to include the \lstinline{Prelude} in 
\ref{prog:h.o.hello_world}, \ref{prog:user_input_client_server}, \ref{prog:broadcastmsgs}, \ref{prog:stack_print}, and \ref{prog:mem-cell}.

\begin{lstlisting}
-- combines code from Examples 14, 31:
protocol Terminal => S =
    StringTerminalPut :: Put([Char]|S) => S
    StringTerminalGet :: Get([Char]|S) => S 
    StringTerminalClose :: TopBot => S
    IntTerminalPut :: Put(Int|S) => S
    IntTerminalGet :: Get(Int|S) => S 
    IntTerminalClose :: TopBot => S
    CharTerminalPut :: Put(Char|S) => S
    CharTerminalGet :: Get(Char|S) => S 
    CharTerminalClose :: TopBot => S

coprotocol S => Console =
    ConsolePut :: S => Get([Char]|S) 
    ConsoleGet :: S => Put([Char]|S) 
    ConsoleClose :: S => TopBot 
    ConsoleStringTerminal :: S => S (*) Neg(Terminal)
    IntConsolePut :: S => Get(Int|S) 
    IntConsoleGet :: S => Put(Int|S) 
    IntConsoleClose :: S => TopBot 
    CharConsolePut :: S => Get(Char|S) 
    CharConsoleGet :: S => Put(Char|S) 
    CharConsoleClose :: S => TopBot 

coprotocol S => Timer =
    Timer :: S => Get(Int|S (*) Put(()|TopBot))
    TimerClose :: S => TopBot

-- from Example 19:
fun isEmpty :: [A] -> Bool =
	[]	-> True
	_		-> False
	
-- from Example 20:
fun (++) :: [A],[A] -> [A] =
	a,[]			-> a
	[],a 			-> a
	(b:bs),cs	-> b : (bs ++ cs)  
\end{lstlisting}

%\newpage
\subsection{Higher order Hello World}
\label{prog:h.o.hello_world}

This program includes the \lstinline{Prelude} from \ref{prog:prelude} and combines code from Examples \ref{ex:helloworld} and \ref{ex:hoprocs}.

\begin{lstlisting}
include Prelude

-- from Example 1:
proc helloworld :: | Console => =
	| console => -> do
		hput ConsolePut on console
		put "Hello World!" on console
		hput ConsoleClose on console
		halt console

-- from Example 18:
proc ho_sender :: | => Put( Store(|Console=>) | TopBot) = 
	| => ch -> do
		on ch do
			put store(helloworld)	
			halt
			
proc ho_receiver :: | Put( Store(|Console=>) | TopBot), Console => =
	| ch, console => -> do
		on ch do
			get stored_process
			close
		on console do
			hput ConsolePut
			put ("Higher order receiver says: Running the stored process")
		use(stored_process)( | console => )

-- new code:
proc run :: | Console => = 
	| console => -> plug
		ho_sender( | => ch)
		ho_receiver( | ch, console => )
\end{lstlisting}

%\newpage
\subsection{Requesting user input continuously}
\label{prog:user_input_client_server}

This program includes the \lstinline{Prelude} from \ref{prog:prelude} and combines code from Examples \ref{ex:protocolpassmsgs}, \ref{ex:serverclientwithservices}, \ref{ex:serverclientwithservices-branching}.

\begin{lstlisting}
include Prelude (isEmpty| )

-- from Example 9:
protocol SendMsgs (A| ) => S =
    SendMsg :: Put(A|S) => S 
    CloseCh :: TopBot => S

-- from Example 24:
proc client :: | => SendMsgs([Char]|), Terminal =
	| => ch, terminal -> do
		on terminal do	
			hput StringTerminalPut
			put "Hello User! Enter message in terminal. Press ENTER to close."
			hput StringTerminalGet
			get msg
		if isEmpty(msg)
			then do
				on ch do 
					hput CloseCh
					close 
				on terminal do
					hput StringTerminalClose	
					halt 
			else do
				on ch do											
					hput SendMsg
					put msg
				client( | => ch, terminal)

-- from Example 15:
proc server :: | SendMsgs([Char]|), Console => =
    | ch, console => -> do
        hcase ch of
            SendMsg -> do
                get msg on ch
                on console do
                    hput ConsolePut
                    put "message from user: " ++ msg
                server( | ch, console => )
            CloseCh -> do
                close ch
                on console do
                    hput ConsoleClose
                    halt
                    
proc run :: | Console => Terminal = 
    | console => terminal -> plug
        client( | => ch, terminal)
        server( | ch, console => )
\end{lstlisting}

%\newpage
\subsection{Broadcasting a single message}
\label{prog:broadcastmsg}

Note that this program will not produce observable effects when it is run.
This program combines code from Examples \ref{ex:broadcastmsg}, \ref{ex:negdest1}, \ref{ex:sender}.
\begin{lstlisting}
-- from Example 28:
proc sender :: | => Put([Char] | Get([Char] | TopBot)) =
	| => source ->
		on source do
			put "Hi everyone!"
			get ack_msg
			halt
			
-- from Example 3:
proc broadcast :: [Char] | Put([Char] | Get([Char] | TopBot)) 								=> Put([Char] | TopBot), Put([Char] | TopBot) =
	ack_msg | source => dest1, dest2 -> do
		get msg on source
		put msg on dest1
		put msg on dest2
		put ack_msg on source
		close source
		close dest1
		halt dest2
		
-- from Example 8:
proc send_to_dest1 :: | Put([Char] | TopBot), Neg(Put([Char] | TopBot)) => =
	| source, neg_dest1 => -> do
		on source do
			get msg
			close
		plug
			neg_dest1, dest1 => ->
				neg_dest1 |=| neg dest1
			=> dest1 ->
				on dest1 do	
					put msg
					halt
					
-- new code:
proc proc1 :: | => Neg(Put([Char] | TopBot)) =
	| => neg_dest1 -> plug
		=> neg_dest1, dest1 ->
			neg_dest1 |=| neg dest1
		dest1 => ->
			on dest1 do	
				get msg
				halt

proc proc2 :: | Put([Char] | TopBot) => =
	| dest2 => ->
		on dest2 do
			get msg
			halt

-- from Example 8:
proc run =
	| => -> plug
		sender( | => source)
		broadcast("message broadcasted" | source => dest1, dest2)
		send_to_dest1( | dest1, neg_dest1 => )
		proc1( | => neg_dest1)	
		proc2( | dest2 =>)
\end{lstlisting} 

%\newpage
\subsection{Broadcasting an arbitrary number of messages}
\label{prog:broadcastmsgs}

This program includes the \lstinline{Prelude} from \ref{prog:prelude} and combines code from Examples \ref{ex:protocolpassmsgs}, \ref{ex:protocolpassmsgswconf}, \ref{ex:recbroadcast}, \ref{ex:coprotocolcopassmsgs},  \ref{ex:coprotocoldest1}.
It also uses ideas from Examples \ref{ex:serverclientwithservices} and \ref{ex:serverclientwithservices-branching}.
We visualize the process network for this program in the following diagram:
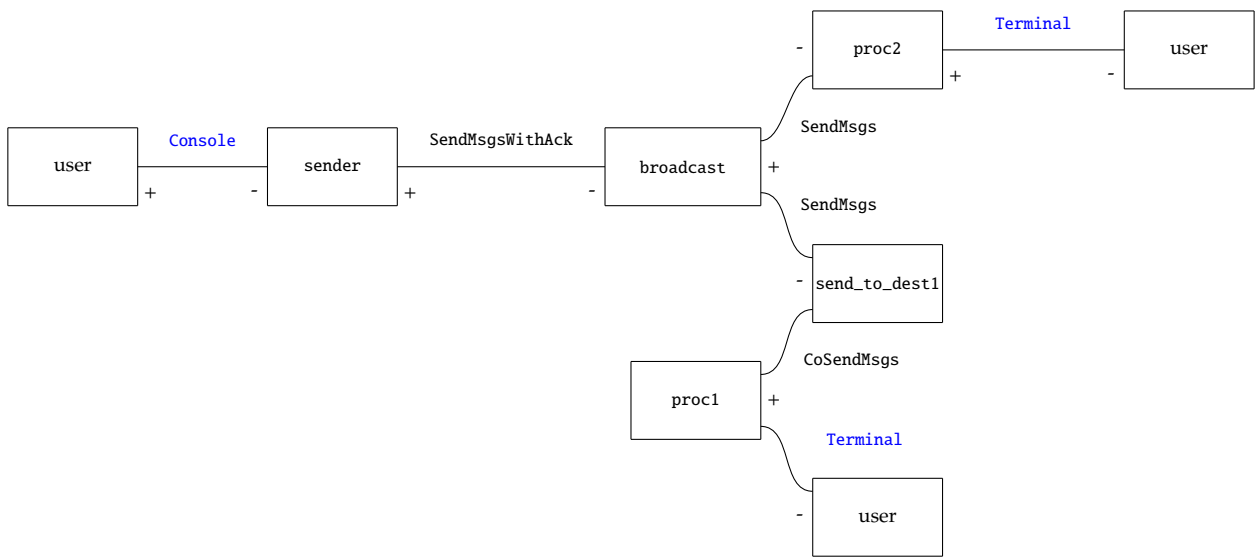
\[
\begin{scaletikzpicturetowidth}{\textwidth}
\begin{tikzpicture}[scale=\tikzscale]
	\begin{pgfonlayer}{nodelayer}
		\node [style=none] (0) at (-16.5, 0.75) {};
		\node [style=none] (1) at (-16.5, -0.75) {};
		\node [style=none] (2) at (-14, -0.75) {};
		\node [style=none] (3) at (-14, 0.75) {};
		\node [style=none] (4) at (-14, 0) {};
		\node [style=none] (5) at (-11.5, -0.75) {};
		\node [style=none] (6) at (-11.5, 0.75) {};
		\node [style=none] (7) at (-11.5, 0) {};
		\node [style=none] (8) at (-9, -0.75) {};
		\node [style=none] (9) at (-9, 0.75) {};
		\node [style=none] (10) at (-9, 0) {};
		\node [style=none] (11) at (-5, -0.75) {};
		\node [style=none] (12) at (-5, 0.75) {};
		\node [style=none] (13) at (-2, -0.75) {};
		\node [style=none] (14) at (-2, 0.75) {};
		\node [style=none] (15) at (-5, 0) {};
		\node [style=none] (16) at (-2, 0.5) {};
		\node [style=none] (17) at (-2, -0.5) {};
		\node [style=none] (18) at (-1, -1.5) {};
		\node [style=none] (19) at (-1, -3) {};
		\node [style=none] (20) at (-1, -1.75) {};
		\node [style=none] (21) at (-1, -2.75) {};
		\node [style=none] (22) at (1.5, -3) {};
		\node [style=none] (23) at (1.5, -1.5) {};
		\node [style=none] (24) at (-2, -3.75) {};
		\node [style=none] (25) at (-2, -5.25) {};
		\node [style=none] (26) at (-2, -4) {};
		\node [style=none] (27) at (-2, -5) {};
		\node [style=none] (28) at (-4.5, -5.25) {};
		\node [style=none] (29) at (-4.5, -3.75) {};
		\node [style=none] (30) at (-1, -6) {};
		\node [style=none] (31) at (-1, -7.5) {};
		\node [style=none] (32) at (-1, -6.25) {};
		\node [style=none] (33) at (1.5, -6) {};
		\node [style=none] (34) at (1.5, -7.5) {};
		\node [style=none] (35) at (-1, 1.5) {};
		\node [style=none] (36) at (-1, 3) {};
		\node [style=none] (37) at (-1, 1.75) {};
		\node [style=none] (38) at (1.5, 1.5) {};
		\node [style=none] (39) at (1.5, 3) {};
		\node [style=none] (40) at (1.5, 2.25) {};
		\node [style=none] (41) at (5, 3) {};
		\node [style=none] (42) at (5, 1.5) {};
		\node [style=none] (43) at (7.5, 1.5) {};
		\node [style=none] (44) at (7.5, 3) {};
		\node [style=none] (45) at (5, 2.25) {};
		\node [style=none] (46) at (-12.75, 0.5) {\lstinline{Console}};
		\node [style=none] (47) at (-7, 0.5) {\lstinline{SendMsgsWithAck}};
		\node [style=none] (48) at (-0.5, 0.75) {\lstinline{SendMsgs}};
		\node [style=none] (49) at (-0.5, -0.75) {\lstinline{SendMsgs}};
		\node [style=none] (50) at (3.25, 2.75) {\lstinline{Terminal}};
		\node [style=none] (51) at (-0.25, -3.75) {\lstinline{CoSendMsgs}};
		\node [style=none] (52) at (0, -5.25) {\lstinline{Terminal}};
		\node [style=none] (53) at (-15.25, 0) {user};
		\node [style=none] (54) at (-10.25, 0) {\lstinline{sender}};
		\node [style=none] (55) at (-3.5, 0) {\lstinline{broadcast}};
		\node [style=none] (56) at (0.25, 2.25) {\lstinline{proc2}};
		\node [style=none] (57) at (6.25, 2.25) {user};
		\node [style=none] (58) at (0.25, -2.25) {\lstinline{send_to_dest1}};
		\node [style=none] (59) at (0.25, -6.75) {user};
		\node [style=none] (60) at (-3.25, -4.5) {\lstinline{proc1}};
		\node [style=none] (61) at (-13.75, -0.5) {+};
		\node [style=none] (62) at (-11.75, -0.5) {-};
		\node [style=none] (63) at (-8.75, -0.5) {+};
		\node [style=none] (64) at (-5.25, -0.5) {-};
		\node [style=none] (65) at (-1.75, 0) {+};
		\node [style=none] (66) at (-1.25, 2.25) {-};
		\node [style=none] (67) at (1.75, 1.75) {+};
		\node [style=none] (68) at (4.75, 1.75) {-};
		\node [style=none] (69) at (-1.25, -2.25) {-};
		\node [style=none] (70) at (-1.75, -4.5) {+};
		\node [style=none] (71) at (-1.25, -6.75) {-};
	\end{pgfonlayer}
	\begin{pgfonlayer}{edgelayer}
		\draw (0.center) to (3.center);
		\draw (1.center) to (2.center);
		\draw (0.center) to (1.center);
		\draw (3.center) to (2.center);
		\draw (6.center) to (9.center);
		\draw (5.center) to (8.center);
		\draw (6.center) to (5.center);
		\draw (9.center) to (8.center);
		\draw (11.center) to (13.center);
		\draw (14.center) to (12.center);
		\draw (12.center) to (11.center);
		\draw (14.center) to (13.center);
		\draw (35.center) to (38.center);
		\draw (39.center) to (38.center);
		\draw (39.center) to (36.center);
		\draw (36.center) to (35.center);
		\draw (18.center) to (23.center);
		\draw (18.center) to (19.center);
		\draw (19.center) to (22.center);
		\draw (23.center) to (22.center);
		\draw (29.center) to (24.center);
		\draw (29.center) to (28.center);
		\draw (28.center) to (25.center);
		\draw (25.center) to (24.center);
		\draw (41.center) to (44.center);
		\draw (41.center) to (42.center);
		\draw (42.center) to (43.center);
		\draw (44.center) to (43.center);
		\draw (30.center) to (33.center);
		\draw (30.center) to (31.center);
		\draw (31.center) to (34.center);
		\draw (33.center) to (34.center);
		\draw (4.center) to (7.center);
		\draw (10.center) to (15.center);
		\draw [in=0, out=-180, looseness=0.75] (37.center) to (16.center);
		\draw [in=-180, out=0] (17.center) to (20.center);
		\draw [in=0, out=-180] (21.center) to (26.center);
		\draw [in=180, out=0] (27.center) to (32.center);
		\draw (40.center) to (45.center);
	\end{pgfonlayer}
\end{tikzpicture}
\end{scaletikzpicturetowidth}
\]

\begin{lstlisting}
include Prelude (isEmpty| )

-- from Example 9:
protocol SendMsgs(A| ) => S =
	SendMsg :: Put(A|S) => S 
	CloseCh :: TopBot => S

-- from Example 10:
protocol SendMsgsWithAck(A| ) => S =
	SendMsgWithAck :: Put(A|Get([Char]|S)) => S
	CloseAckCh :: TopBot => S 
	
-- from Example 12:
coprotocol S => CoSendMsgs(A| ) =
	CoSendMsg ::  S => Get(A|S)
	CoCloseCh :: S => TopBot

-- new code (combines ideas from Examples 15 and 24):
proc sender :: | Console => SendMsgsWithAck([Char]| ) =
	| console => source -> do
		on console do	
			hput ConsolePut
			put "Hello User! Enter message to broadcast. Press ENTER to close."
			hput ConsoleGet
			get msg
		if isEmpty(msg)
			then do
				on console do
					hput ConsolePut
					put "Indicating to destination processes that source is finished."
				on source do 
					hput CloseAckCh
					close 
				on console do
					hput ConsoleClose	
					halt 
			else do
				on source do											
					hput SendMsgWithAck
					put msg
					get ack_msg
				on console do	
					hput ConsolePut
					put ack_msg
				sender( | console => source)

-- from Example 11:
proc broadcast :: [Char] | SendMsgsWithAck([Char]| ) => SendMsgs([Char]| ), SendMsgs([Char]| ) = 
	ack_msg | source => dest1, dest2 -> do
		hcase source of
			SendMsgWithAck -> do
				get msg on source
				on dest1 do
					hput SendMsg
					put msg
				on dest2 do
					hput SendMsg
					put msg
				on source do
					put ack_msg
				broadcast(ack_msg | source => dest1, dest2)
			CloseAckCh -> do
				close source							
				on dest1 do						
					hput CloseCh
					close
				on dest2 do
					hput CloseCh	
					halt

-- from Example 13:
proc send_to_dest1 :: | SendMsgs([Char]| ), CoSendMsgs([Char]| ) => =
	| source, dest1 => -> 
		hcase source of
			SendMsg -> do
				get msg on source 		
				on dest1 do
					hput CoSendMsg
					put msg
				send_to_dest1( | source, dest1 => )
			CloseCh -> do
				close source
				on dest1 do
					hput CoCloseCh
					halt

-- new code (combines ideas from Examples 15 and 24):
proc proc1 :: [Char] | => CoSendMsgs([Char]| ), Terminal =
	tag | => source, terminal -> do
		hcase source of
			CoSendMsg -> do
				get msg on source 		
				on terminal do
					hput StringTerminalPut
					put tag ++ " received message: " ++ msg
				proc1(tag | => source, terminal)
			CoCloseCh -> do
				close source
				on terminal do
					hput StringTerminalPut
					put "Source has finished sending messages to " ++ tag ++ ". Press ENTER to close."
					hput StringTerminalGet
					get _
					hput StringTerminalClose
					halt

proc proc2 :: [Char] | SendMsgs([Char]| ) => Terminal =
	tag | source => terminal -> do
		hcase source of
			SendMsg -> do
				get msg on source 		
				on terminal do
					hput StringTerminalPut
					put tag ++ " received message: " ++ msg
				proc2(tag | source => terminal)
			CloseCh -> do
				close source
				on terminal do
					hput StringTerminalPut
					put "Source has finished sending messages to " ++ tag ++ ". Press ENTER to close."
					hput StringTerminalGet
					get _
					hput StringTerminalClose
					halt

proc run =
    | console => term1, term2 -> plug
        sender( | console => source)
        broadcast("Message has been broadcasted." | source => dest1, dest2)
        send_to_dest1( | dest1, codest1 => )
        proc1("Process 1" | => codest1, term1)
        proc2("Process 2" | dest2 => term2)
\end{lstlisting}

%\newpage
\subsection{Non-deterministic server interacts with two clients}
\label{prog:non_det_server}

Note that this program will not produce observable effects when it is run.
This program combines code from Examples \ref{ex:serverecho}, \ref{ex:clientfork}, \ref{ex:nondetserver}.

\begin{lstlisting}
-- from Example 16:
defn
	proc server =
		| two_ch => -> do
			split two_ch into ch1, ch2
			server_non_deterministic( | ch1, ch2 => )

where defn
	proc server_non_deterministic =
		| ch1, ch2 => -> do
			race
				ch1 -> server_deterministic( | ch1, ch2 => )	
				ch2 -> server_deterministic( | ch2, ch1 => )	
	
	proc server_deterministic =
		| winner, loser => -> do
			on winner do
				get msg
				put msg
				close
			on loser do
				get msg
				put msg
				halt

-- from Example 2:
proc client = 
	| => ch ->
		on ch do 
			put "Hello Server!"
			get echo	
			halt

-- from Example 7:
proc two_clients =
	| => two_ch ->
		fork two_ch as	
			ch1 -> client( | => ch1)
			ch2 -> client( | => ch2)
			
proc run :: | => = 
 	| => -> plug						
 		two_clients( | => two_ch)	
 		server( | two_ch => )
\end{lstlisting}

%\newpage
\subsection{Pushing messages onto a Stack codata type}
\label{prog:stack_print}

This program includes the \lstinline{Prelude} from \ref{prog:prelude} and combines code from Examples \ref{ex:protocolpassmsgs}, \ref{ex:codata stack}, \ref{ex:record}, \ref{ex:codata reverse_print}, and \ref{ex:serverclientwithservices-branching}.

\begin{lstlisting}
include Prelude (isEmpty| )

-- from Example 9:
protocol SendMsgs(A| ) => S =
	SendMsg :: Put(A|S) => S 
	CloseCh :: TopBot => S

-- from Example 21:
data SF(A) -> C =
	SS :: A -> C
	FF :: -> C
	
codata S -> Stack(A) = 
	Push :: A, S -> S
	Pop :: S -> (SF(A),S)	

-- from Example 22:	
fun listStack :: [A] -> Stack(A) = 
	cs -> (
		Push := c -> listStack(c:cs),
		Pop := -> case cs of	
			b:bs -> (SS(b), listStack(bs))
			[] -> (FF, listStack([]))
		)

-- from Example 23:
defn        
	proc reverse_print :: Stack([Char]) | SendMsgs([Char]| ), Console => =
		stack | source, console => -> do
			hcase source of
				SendMsg -> do
					get msg on source
					reverse_print(Push(msg, stack) | source, console => )
				CloseCh -> do
					close source
					on console do
						hput ConsolePut
						put "Printing stack from top to bottom: " ++ flatten(stack)
						hput ConsoleClose
						halt
where
	fun flatten :: Stack([Char]) -> [Char] =
		s -> case Pop(s) of
			(SS(str), stack) -> str ++ " " ++ flatten(stack)
			(FF, stack) -> ""


-- from Example 24:
proc client :: | => SendMsgs([Char]|), Terminal =
	| => ch, terminal -> do
		on terminal do	
			hput StringTerminalPut
			put "Hello User! Enter message in terminal. Press ENTER to close."
			hput StringTerminalGet
			get msg
		if isEmpty(msg)
			then do
				on ch do 
					hput CloseCh
					close 
				on terminal do
					hput StringTerminalClose	
					halt 
			else do
				on ch do											
					hput SendMsg
					put msg
				client( | => ch, terminal)

-- new code:
proc server :: | SendMsgs([Char]| ), Console => =
	| ch, console => -> do
		on console do
			hput ConsolePut
			put "Collecting messages to print in reverse."
		reverse_print(listStack([]) | ch, console => )	-- initializes stack

proc run :: | Console => Terminal = 
    | console => terminal -> plug
        client( | => ch, terminal)
        server( | ch, console => )
\end{lstlisting}

%\newpage
\subsection{Memory cell}
\label{prog:mem-cell}

This program includes the \lstinline{Prelude} from \ref{prog:prelude} and modifies code from Examples \ref{ex:mutex-protocol} and \ref{ex:mem-cell}.
We visualize the process network for this program in the following diagram:
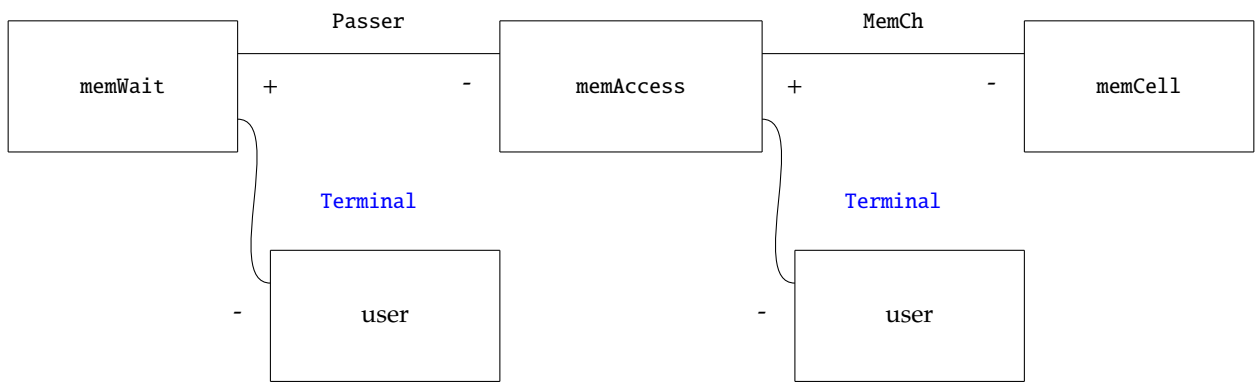
\[
\begin{scaletikzpicturetowidth}{\textwidth}
\begin{tikzpicture}[scale=\tikzscale]
	\begin{pgfonlayer}{nodelayer}
		\node [style=none] (0) at (-7, 1) {};
		\node [style=none] (1) at (-7, -1) {};
		\node [style=none] (2) at (-3.5, 1) {};
		\node [style=none] (3) at (-3.5, -1) {};
		\node [style=none] (4) at (-3.5, -0.5) {};
		\node [style=none] (5) at (-3.5, 0.5) {};
		\node [style=none] (6) at (-3, -2.5) {};
		\node [style=none] (7) at (-3, -4.5) {};
		\node [style=none] (8) at (0.5, -2.5) {};
		\node [style=none] (9) at (0.5, -4.5) {};
		\node [style=none] (10) at (0.5, -1) {};
		\node [style=none] (11) at (0.5, 1) {};
		\node [style=none] (12) at (0.5, 0.5) {};
		\node [style=none] (13) at (4.5, 1) {};
		\node [style=none] (14) at (4.5, -1) {};
		\node [style=none] (16) at (-3, -3) {};
		\node [style=none] (17) at (4.5, 0.5) {};
		\node [style=none] (18) at (4.5, -0.5) {};
		\node [style=none] (19) at (5, -2.5) {};
		\node [style=none] (20) at (5, -4.5) {};
		\node [style=none] (21) at (8.5, -2.5) {};
		\node [style=none] (22) at (8.5, -4.5) {};
		\node [style=none] (23) at (8.5, 1) {};
		\node [style=none] (24) at (8.5, -1) {};
		\node [style=none] (25) at (8.5, 0.5) {};
		\node [style=none] (26) at (12, 1) {};
		\node [style=none] (27) at (12, -1) {};
		\node [style=none] (28) at (5, -3) {};
		\node [style=none] (29) at (-5.25, 0) {\lstinline{memWait}};
		\node [style=none] (30) at (-1.25, -3.5) {user};
		\node [style=none] (31) at (2.5, 0) {\lstinline{memAccess}};
		\node [style=none] (32) at (6.75, -3.5) {user};
		\node [style=none] (33) at (10.25, 0) {\lstinline{memCell}};
		\node [style=none] (34) at (-1.5, 1) {\lstinline{Passer}};
		\node [style=none] (35) at (6.5, 1) {\lstinline{MemCh}};
		\node [style=none] (36) at (-1.5, -1.75) {\lstinline{Terminal}};
		\node [style=none] (40) at (6.5, -1.75) {\lstinline{Terminal}};
		\node [style=none] (41) at (-3, 0) {+};
		\node [style=none] (42) at (0, 0) {-};
		\node [style=none] (43) at (5, 0) {+};
		\node [style=none] (44) at (8, 0) {-};
		\node [style=none] (45) at (4.5, -3.5) {-};
		\node [style=none] (46) at (-3.5, -3.5) {-};
	\end{pgfonlayer}
	\begin{pgfonlayer}{edgelayer}
		\draw (0.center) to (2.center);
		\draw (0.center) to (1.center);
		\draw (1.center) to (3.center);
		\draw (2.center) to (3.center);
		\draw (6.center) to (7.center);
		\draw (7.center) to (9.center);
		\draw (8.center) to (9.center);
		\draw (6.center) to (8.center);
		\draw (11.center) to (13.center);
		\draw (11.center) to (10.center);
		\draw (10.center) to (14.center);
		\draw (13.center) to (14.center);
		\draw (19.center) to (20.center);
		\draw (20.center) to (22.center);
		\draw (19.center) to (21.center);
		\draw (21.center) to (22.center);
		\draw (23.center) to (24.center);
		\draw (24.center) to (27.center);
		\draw (26.center) to (27.center);
		\draw (23.center) to (26.center);
		\draw [in=180, out=0, looseness=0.75] (4.center) to (16.center);
		\draw (5.center) to (12.center);
		\draw [in=-180, out=0, looseness=0.75] (18.center) to (28.center);
		\draw (17.center) to (25.center);
	\end{pgfonlayer}
\end{tikzpicture}
\end{scaletikzpicturetowidth}
\]

\begin{lstlisting}
include Prelude (isEmpty| )
    
-- modified from Example 26:
protocol Passer( |R) => S =
	Pass :: R (+) Neg(S) => S
	Done :: TopBot => S 									-- this handle is new
    
-- new code:
protocol MemCh(A| ) => S =							-- memory cell access protocol
	MemPut :: Put(A|S) => S 							-- write
	MemGet :: Get(A|S) => S 							-- read
	MemCls :: TopBot => S 								-- close

proc memCell :: A | MemCh(A| ) => =			-- memory cell process
	val | ch => -> hcase ch of
		MemPut -> do 												-- overwrite existing stored value
			get nval on ch
			memCell(nval | ch => )
		MemGet -> do 												-- send stored value
			put val on ch
			memCell(val | ch => )
		MemCls -> do 												-- close
			halt ch

proc memDone :: | Passer( |MemCh([Char]| )) => MemCh([Char]| ),  Terminal =
	| passer => mem, terminal -> do
		on terminal do 											-- user wants to close
			hput StringTerminalClose
			close
		hcase passer of  										-- check if other proc wants mem
			Pass -> do
				fork passer as
					pass_mem with mem -> 					-- if they do, pass it back
						pass_mem |=| mem
					neg_passer -> plug
						neg_passer, new_passer => ->		
							neg_passer |=| neg new_passer
						=> new_passer -> on new_passer do
							hput Done									-- but don't request it again
							halt
			Done -> do 												-- close if other proc is already done
				close passer
				on mem do
					hput MemCls
					halt
					
-- modified from Example 27:
defn
	proc memAccess :: [Char] | Passer( |MemCh([Char]| )) => MemCh([Char]| ), Terminal =
		tag | passer => mem, terminal -> do	-- terminal is new
			on mem do
				hput MemGet
				get mem_data
			on terminal do										-- print mem data, get user input
				hput StringTerminalPut
				put tag ++ " read value: " ++ mem_data
				hput StringTerminalPut
				put tag ++ ", please enter a string or press ENTER to close."
				hput StringTerminalGet
				get user_input
			if isEmpty(user_input)						-- check if user wants to close
				then memDone( | passer => mem, terminal)
				else do													-- if they don't, it's the same
					on mem do
						hput MemPut
						put user_input
					hcase passer of 
						Pass -> do
							fork passer as
								pass_mem with mem -> 
									pass_mem |=| mem
								neg_passer with terminal -> plug
									neg_passer, new_passer => ->		
										neg_passer |=| neg new_passer
									memWait(tag | => new_passer, terminal)	-- and keep terminal
						Done -> do									-- but other proc might be done
							close passer
							on mem do
								hput MemCls
								close
							on terminal do
								hput StringTerminalPut
								put "Other process is done. Press ENTER to close."
								hput StringTerminalGet
								get _
								hput StringTerminalClose
								halt
		
	proc memWait :: [Char] | => Passer( |MemCh([Char]| )), Terminal =
		tag | => passer, terminal -> do										-- terminal is new
			hput Pass on passer
			split passer into mem, neg_passer
			plug
				=> neg_passer, new_passer -> 
					neg_passer |=| neg new_passer
				memAccess(tag | new_passer => mem, terminal)	-- and used in memAccess

proc run =
	| => terminalPing, terminalPong -> plug 
		memCell( "" | mem => )
		memAccess("Ping" | passer => mem, terminalPing)
		memWait("Pong" | => passer, terminalPong)
\end{lstlisting}

\end{appendix}

\end{document}